\documentclass[onefignum,onetabnum]{siamonline250106}

\usepackage{amsmath}
\usepackage{amsfonts}
\usepackage{graphicx}
\usepackage{subcaption}
\usepackage{xcolor}
\usepackage{pdflscape}
\usepackage{algorithm}
\usepackage{algpseudocode}
\usepackage{float}

\ifpdf
  \DeclareGraphicsExtensions{.eps,.pdf,.png,.jpg}
\else
  \DeclareGraphicsExtensions{.eps}
\fi

\usepackage{enumitem}
\setlist[enumerate]{leftmargin=.5in}
\setlist[itemize]{leftmargin=.5in}

\usepackage{placeins}

\algnewcommand\algorithmicinput{\textbf{Input:}}
\algnewcommand\algorithmicoutput{\textbf{Output:}}
\algnewcommand\Input{\item[\algorithmicinput]}
\algnewcommand\Output{\item[\algorithmicoutput]}
\DeclareMathOperator{\nullspace}{null}
\DeclareMathOperator{\rowspace}{row}
\DeclareMathOperator{\Span}{span}
\DeclareMathOperator{\rank}{rank}

\headers{Invariant Image Reparameterisation}{O. J. Maclaren et al.}

\usepackage{hyperref}

\ifpdf
\hypersetup{
  pdftitle={Invariant Image Reparameterisation},
  pdfauthor={O. J. Maclaren, R. Nicholson, J. A. Trent, J. Rottenberry, and M. J. Simpson}
}
\fi

\title{Invariant Image Reparameterisation: Bridging Symbolic and Numerical Methods for Identifiability Analysis, Model Reduction, and Prediction\thanks{This is the accepted manuscript of an article published in \emph{Mathematical Biosciences} 403 (2027) 109844, \url{https://doi.org/10.1016/j.mbs.2026.109844}. \copyright~2026 The Authors. This manuscript version is made available under the CC BY 4.0 license \url{https://creativecommons.org/licenses/by/4.0/}.}}

\author{%
  Oliver~J.~Maclaren\thanks{Department of Engineering Science and Biomedical Engineering, University of Auckland, Auckland 1142, New Zealand (\email{oliver.maclaren@auckland.ac.nz}).} \and
  Ruanui Nicholson\footnotemark[2] \and
  Joel~A.~Trent\footnotemark[2] \and
  Joshua Rottenberry\thanks{School of Mathematical Sciences, Queensland University of Technology (QUT), Brisbane, Australia.} \and
  Matthew~J.~Simpson\footnotemark[3]\thanks{ARC Centre of Excellence for the Mathematical Analysis of Cellular Systems, QUT, Brisbane, Australia.}%
}

\date{}

\begin{document}

\maketitle

\begin{abstract}
Structural and practical parameter non-identifiability issues are common when mathematical models are used to interpret data. Such issues motivate model reparameterisation and reduction methods. Here, we consider Invariant Image Reparameterisation (IIR), which asks when symbolic reparameterisation conditions can be replaced by numerical derivative calculations at a single reference point. The central object is the invariant image: a reduced, basis-independent representation of the parameter combinations controlling observable model behaviour. We show that when a one-to-one componentwise transformation makes observable behaviour depend only on fixed linear combinations of the transformed parameters, a single numerical Jacobian determines the associated lower-dimensional reparameterisation space. This includes models depending on monomial combinations of the original parameters. We also give a first-order invariance condition that can rule out a minimal reduction of this form and extracts candidate invariant null directions for an exact non-minimal reduction. In structurally identifiable but practically weakly informed settings, the same calculations separate strongly and weakly informed parameter combinations. The invariant image admits multiple coordinate representations: the SVD gives a default orthonormal basis ordered by local identifiability, while sparse monomial bases are often more interpretable. Treating these coordinates as interest parameters in Profile-Wise Analysis gives likelihood-based uncertainty quantification and prediction. We demonstrate the method on parameterised normal models with Poisson-limit, extended Poisson-limit, and non-limit cases, and on the repressilator, a nonlinear differential equation model of gene regulation. A Julia implementation of IIR, with these and further examples, is available at \href{https://github.com/omaclaren/reparam}{\texttt{https://github.com/omaclaren/reparam}}.
\end{abstract}


\section{Introduction}
Mechanistic mathematical models are widely used in mathematical biology and related areas to interpret experimental and observational data, infer underlying mechanisms, and make predictions about observed and unobserved quantities. A central challenge is determining whether the available observations uniquely identify model parameters or, at least, provide sufficient precision for mechanistic interpretation or informed prediction~\cite{Simpson2023,Simpson2024,Villaverde2023}. \textit{Structural identifiability analysis}~\cite{Bellu2007, Chis2016, Ligon2018, Meshkat2009, Meshkat2015, Maclaren2019, Diaz-Seoane2022} asks whether model parameters are uniquely determined by ideal, noise-free observations of chosen model outputs. \textit{Practical identifiability analysis}~\cite{Hines2014, Maclaren2019, Wieland2021, Kreutz2012} asks whether a finite amount of noisy data allows sufficiently precise parameter estimates. Structural non-identifiability occurs when distinct parameter values generate precisely the same observable behaviour, leading to non-uniqueness in inversion; practical non-identifiability occurs when different parameter values generate outputs too similar to be reliably distinguished given the noise level and available outputs, leading to instability or lack of precision in inversion. Identifiability issues are not the end of model analysis, however. When the full model parameter vector is not identifiable it can be possible to determine identifiable parameter combinations~\cite{Maclaren2019, Cole2020}. We may also be able to perform model reduction to obtain simpler, potentially identifiable models~\cite{Catchpole1998, Chappell1998, Evans2000, Meshkat2009, Meshkat2015, Maiwald2016, Cole2020}.

Existing identifiability analysis approaches address different aspects of identifiability problems and goals. Symbolic methods~\cite{Catchpole1998, Chappell1998, Cole2010, Chis2011} can yield exact structural identifiability results but typically only apply in idealised settings; Fisher information~\cite{Rothenberg1971,Cole2020}, sensitivity-based methods, active subspaces~\cite{Constantine2016,Brouwer2018}, and likelihood-informed dimension reduction~\cite{Cui2014,Cui2022} can all guide parameter-space dimension reduction, using local derivative or curvature information either at a reference point or averaged over a chosen distribution. These approaches scale better than symbolic methods, but their resulting subspaces are generally approximate and do not usually provide globally valid symbolic reparameterisations. Profile likelihood methods~\cite{Kreutz2013, Raue2014, Maiwald2016, Simpson2023} provide powerful practical identifiability diagnostics but usually require the user to choose target parameters or subsets in advance, or to perform significant manual analysis to guide reduction~\cite{Eisenberg2014, Maiwald2016}. Sloppiness and related approaches~\cite{Brown2003, Gutenkunst2007, Monsalve2022} can highlight locally dominant and weak directions but are closely related to, and share similar limitations with, Fisher information and sensitivity-based methods, and often rely on non-parameterisation-invariant approximations for uncertainty quantification. Furthermore, in applications of sloppiness and profile likelihood analysis to model reduction, users often simply set poorly identifiable parameters to zero or some arbitrary value~\cite{Lawrie2007, Elevitch2020, Vollert2023}. This can lead to setting individually poorly identifiable parameters to values that jointly violate the requirements on possible values for the identifiable parameter combinations.

As described in the Methods, the symbolic reparameterisation condition used as the starting point for the present work is essentially equivalent to the approach in~\cite{Catchpole1997,Catchpole1998}, summarised and extended in~\cite{Cole2010,Cole2020}. We build on this to consider when derivative information computed numerically at a single point can provide the same reparameterisation information as such symbolic calculations. This connects to a central gap in the literature: understanding when local numerical calculations can recover parameter combinations that remain valid beyond the local approximation region, without requiring further symbolic calculation to establish their validity.

Two other lines of work are closely related to the present approach. The first is that of Joubert, Stigter, Molenaar, Van Willigenburg, and co-workers~\cite{Joubert2020,VanWilligenburg2022}. Like ours, this line of work uses local derivative information and singular value decompositions to diagnose local non-identifiability and find parameter relationships in non-identifiable models. In contrast to our work, Joubert et al.~\cite{Joubert2020} use initial numerical sensitivity calculations to guide subsequent reparameterisation using explicit symbolic calculation, whereas here we ask when local numerical calculation alone can provide the symbolic reparameterisation information. The second related work is the subset-profiling approach of Eisenberg and Hayashi~\cite{Eisenberg2014}. This approach uses numerical calculations of the local Fisher information matrix and profile likelihoods over subsets of parameters to determine potential parameter relationships. The final relationships in this approach are obtained by numerically fitting functions to the profile relationships, with emphasis on rational functions. This aligns with our emphasis on numerical calculation, but the function-fitting step, in general, does not establish that the recovered relationships are exact, globally valid parameter-combination relationships.

We show that when a one-to-one, componentwise transformation of the parameters makes the observable behaviour depend only on fixed lower-dimensional linear combinations of the transformed parameters, and the reduced map has an injective derivative at a reference point, the symbolic reparameterisation condition has a local numerical form: a Jacobian calculation at that point determines a reduced parameter space representing the parameter combinations on which the observable model behaviour depends globally. We call this reduced space the invariant image. While symbolic approaches usually focus on minimal identifiable reparameterisations, we also consider non-minimal image reparameterisations that provide partial reductions. These partial reductions can give valid lower-dimensional representations of the observable behaviour even when further reduction is possible. A first-order invariance condition can rule out the minimal case and extracts candidate invariant null directions for an exact non-minimal reduction. The same calculations can separate strongly and weakly informed parameter combinations in structurally identifiable but practically weakly informed models.

The reduced space can be parameterised in different ways. Different parameter combinations can be understood as coordinate choices within the same invariant image; we therefore consider different possible bases for representing it. A default orthonormal basis is obtained from the SVD, with directions ordered by local identifiability, but this basis may be dense and difficult to interpret. We also consider sparse monomial bases, which are typically more interpretable. Coordinates chosen within the invariant image support model reduction and provide natural parameters of interest for our Profile-Wise Analysis (PWA) framework~\cite{Simpson2024, Simpson2023}, yielding targeted interval estimates for both interest parameters and predictions.

To illustrate these ideas, we first apply our method to a parameterised normal model with practical identifiability issues and a limiting structurally non-identifiable model, before considering a more complex mechanistic model of gene regulation, the repressilator~\cite{Elowitz2000,Eisenberg2014}. The repressilator provides a simple case in which the model structure, without additional transformations, falls outside the standard polynomial/rational form required by some differential-algebra approaches to symbolic identifiability analysis, while the model outputs and derivatives remain numerically accessible. Additional examples are available in the accompanying repository and summarised briefly in the Supplementary Material.

\section{Methods}
Here, we outline the technical foundations of our `Invariant Image Reparameterisation' (IIR) approach. This builds on ideas in~\cite{Catchpole1997,Catchpole1998,Cole2010,Cole2020,Simpson2023}, and, indirectly,~\cite{Maclaren2019, Maclaren2021}.

We present the method focusing on the relation between symbolic and numerical reparameterisation calculations. First, we state an exact symbolic reparameterisation condition in terms of differential conditions on the row and null spaces of the auxiliary mapping Jacobian. Second, we show that, for reparameterisations that can be written as a sequence of transformations consisting of a componentwise change of variables, a constant linear reduction in the transformed variables, and an optional componentwise reparameterisation of the reduced coordinates, the symbolic condition can be represented exactly by local derivative information at a single reference point, provided the reduced map has an injective derivative there. This representation makes the reparameterisation problem accessible to numerical calculation. Finally, we describe a first-order invariance calculation that can rule out the minimal case and extracts candidate invariant directions in the local null space for exact non-minimal reductions. We also explain how the same machinery can be used in approximate, practical settings.

\subsection{Model representation: auxiliary mapping}
With mechanistic models, it is useful to distinguish between underlying theoretical model parameters and parameters more directly associated with observable data or data distributions. We use $\theta \in \Theta \subseteq \mathbb{R}^p$ to denote the model parameters, and $y \in \mathcal{Y} \subseteq \mathbb{R}^n$ to denote observed or predicted data. We assume a statistical model for the data, $p(y;\phi)$, where $\phi \in \Phi \subseteq \mathbb{R}^q$ denotes data distribution parameters. These may characterise a statistical observation model on top of a deterministic mechanistic model, but more generally represent any statistical model capturing key features of the data. This is compatible with the underlying generating model being stochastic.

The role of the mechanistic model is then to connect $\theta$ to these data distribution parameters. We call the resulting deterministic mapping of parameters to parameters the auxiliary mapping $\phi(\theta)$~\cite{Simpson2023}, which may be explicit or may require numerical computation. This gives
\begin{equation}
\theta \mapsto \phi(\theta),
\end{equation}
and, combined with the statistical model,
\begin{equation}
p(y;\theta) = p(y;\phi(\theta)).
\end{equation}

This general approach has been used in econometrics, for example in indirect inference and the method of moments~\cite{Gourieroux1993,Heggland2004}, and in mechanistic modelling in biology and engineering~\cite{Drovandi2015,Browning2020,Simpson2021,Simpson2023}. There are also close connections to the `exhaustive summary' approach reviewed by~\cite{Cole2010,Cole2020}, where our data distribution parameters correspond to the exhaustive summaries. A simple example occurs in the method of moments, where the data distribution parameters are the mean and variance, while the underlying model parameters may be more complex.

As an illustrative running example, we use the normal approximation $Y\sim\mathcal{N}(np,np)$ to a binomial model near its Poisson limit, with positive parameters $n$ and $p$; this example is defined in Section~\ref{sec:Models} and analysed fully in the Results. Its auxiliary mapping gives the mean and variance, $\phi(n,p)=(np,np)^T$.

For deterministic spatial or temporal models considered here, $\phi(\theta)$ may be a numerical solution on a fine grid, or a collection of model predictions at specified times or locations. Although our main mechanistic example involves ordinary differential equations in time, identifiability questions for partial differential equation models are also important in mathematical biology~\cite{Renardy2022, Browning2024}. The auxiliary-map approach we use here requires only a numerically computable mapping from parameters to observable quantities, together with numerically computable derivatives of this mapping. Hence, it can in principle be applied to numerical solutions of partial differential equations or boundary-value ordinary differential equations. An example of the latter is considered in the repository examples. Similarly, for a stochastic model, one might use the full trajectories of moments defined by a moment dynamics approximation~\cite{Browning2020}. In our examples, the required derivatives are evaluated by automatic differentiation in Julia~\cite{Revels2016}. This is distinct from symbolic differentiation: automatic differentiation at the source-code level evaluates the numerical value of the derivative at a fixed input value~\cite{Griewank2008}. Numerical solutions and their derivatives are typically more readily available for complex models than symbolic expressions.

\subsection{Image reparameterisations of the auxiliary mapping}
A key idea in identifiability analysis is that non-identifiable parameters can often be combined into identifiable parameter combinations or, more generally, that some functions of the parameters can be made identifiable~\cite{Maclaren2019,Cole2020}. In terms of the auxiliary mapping, a reduced-dimension reparameterisation corresponds to decomposing
\begin{equation}
\phi(\theta) = \tilde{\phi}(\psi(\theta)),
\label{eq:reparam}
\end{equation}
where $\psi(\theta)$ gives a lower-dimensional parameterisation of the observable behaviour, and $\tilde{\phi}$ maps this reduced parameterisation to the data distribution parameters. This decomposition is illustrated later in Figure~\ref{fig:epi-linear}, where it is shown together with the transformed-coordinate construction used by IIR.

Although parameter combinations are a natural target, they are not generally uniquely determined, even when the parameterisation is reduced to an identifiable subset. This is true of both symbolic and numerical approaches~\cite{Cole2010,Cole2020}, as even identifiable models are usually invariant under one-to-one reparameterisation. Thus, the primary object determined by our approach is what we call the invariant image associated with the auxiliary mapping, rather than any single set of parameter combinations. This image is basis-independent, while different reparameterisations correspond to different coordinate choices within it.

It is natural to ask whether there exists a factorisation of the form \eqref{eq:reparam} for which $\psi(\theta)$ extracts identifiable parameter combinations and $\tilde{\phi}$ is one-to-one. This gives what we call a minimal image reparameterisation. We also allow partial reductions, where the dimension is reduced but redundancy is not fully removed. We call these non-minimal image reparameterisations. Such reductions can still be useful as reduced representations in their own right, or as intermediate steps towards a minimal reparameterisation. In the running example, $\psi(n,p)=np$ and $\tilde{\phi}(\psi)=(\psi,\psi)^T$ give a minimal image reparameterisation, since $\tilde{\phi}$ is one-to-one in $\psi$.

The existence of a minimal image reparameterisation is guaranteed abstractly by the fact that any mapping can be decomposed into an onto mapping followed by a one-to-one mapping (the epi--mono factorisation of the category of sets and mappings~\cite{Lawvere2003}). However, this abstract result does not provide a practical method for finding the intermediate image or associated reparameterisation.

The rest of this section concerns making this concept practical, first symbolically and then numerically. For this, we assume our mappings are sufficiently smooth and use the language of differential geometry relatively informally~\cite{Lee2013}. The clearest class of models, and the class we mainly focus on here, is one in which the model can be expressed in terms of monomial parameter combinations. This is a natural assumption for many mechanistic models, for example via dimensional analysis~\cite{Buckingham1914,Barenblatt2003}, and also arises in asymptotic approximations and model reduction~\cite{Maiwald2016}. For positive parameters, or parameters of known sign that can be written in terms of positive ones, taking logarithms turns monomial combinations, such as products, ratios and powers, into linear combinations of log parameters. This motivates our focus on logarithmic parameter coordinates. However, the framework is not restricted to monomial combinations: other componentwise transformations and coordinate choices can lead to other classes of reparameterisations, and non-minimal image reparameterisations can provide intermediate spaces in which further reductions can be tested. The componentwise assumption is also not strictly essential to the chain-rule argument but is convenient and interpretable for the present implementation. Geometrically, the ideas here have strong connections to symmetry-based approaches to structural identifiability and, in the log-monomial case, to scaling-based methods~\cite{Massonis2020,Massonis2023,Castro2020,Villaverde2021}. Our focus here is on when such reparameterisations can be recovered numerically from local auxiliary-map derivatives.

\subsection{Symbolic reparameterisation condition}
Given an arbitrary reparameterisation as in \eqref{eq:reparam}, we further assume we are working with smooth mappings with constant rank. Applying the chain rule then gives a differential version of the reparameterisation equation:
\begin{equation}
D_\theta\phi(\theta) = D_\psi\tilde{\phi}(\psi(\theta))D_\theta\psi(\theta).
\label{eq:diff-reparam}
\end{equation}
This implies
\begin{equation}
    \begin{aligned}
\nullspace D_\theta\psi(\theta) &\subseteq \nullspace D_\theta\phi(\theta)\\
\rowspace D_\theta\psi(\theta) &\supseteq \rowspace D_\theta\phi(\theta).
\end{aligned}
\end{equation}
This is a simple statement of linear algebra but provides useful intuition and points to constructive algorithms for reparameterisation. Firstly, this shows that the redundancy, i.e., reduction in dimension, associated with a (possibly non-minimal) reparameterisation is less than or equal to the overall redundancy, i.e., reduction in dimension, associated with the full parameter-to-data space mapping. Conversely for the preserved dimension components associated with the row space. Secondly, it provides a method for finding a reduced dimension parameterisation. Since the goal is usually, though not always, a full reduction to a minimal reparameterisation, we discuss this first in the context of minimal reparameterisations.

Analogously to minimal reparameterisations in the abstract case, in the differential case we strengthen the assumptions about the mappings involved by requiring that $D_\psi\tilde{\phi}(\psi(\theta))$ be one-to-one. This amounts to strengthening the assumptions about the original minimal image reparameterisation (epi-mono factorisation) to requiring $\psi(\theta)$ to be a smooth submersion and $\tilde{\phi}(\psi)$ to be a smooth immersion~\cite{Lee2013}. Now, since $D_\psi\tilde{\phi}(\psi(\theta))$ is one-to-one, the subset relations are strengthened to equalities and $D_\theta\psi(\theta)$ and $D_\theta\phi(\theta)$ share exactly the same null space/kernel and complementary row spaces:
\begin{equation}
\begin{aligned}
\nullspace D_\theta\psi(\theta) &= \nullspace D_\theta\phi(\theta)\\
\rowspace D_\theta\psi(\theta) &= \rowspace D_\theta\phi(\theta).
\end{aligned}
\end{equation}
Next, constructively, given null-space vectors $\alpha_j(\theta)$ obtained from the full model derivative $D_\theta\phi(\theta)$, we can determine a reparameterisation by solving the differential condition:
\begin{equation}
D_\theta\psi(\theta)\alpha_j(\theta) = 0,\qquad j=1,\dots,p-r,
\end{equation}
for $\psi(\theta)$. Solving these equations symbolically gives a candidate reparameterisation. To obtain a minimal image reparameterisation, a full set of null-space vectors of $D_\theta\phi(\theta)$ must be used in these equations, and $D_\theta\psi(\theta)$ should have no additional null directions (equivalently, the row space is preserved as in the subspace conditions above). In the running example, $D_\theta\phi$ has null vector $\alpha=(n,-p)^T$, so the condition is $n\,\partial_n\psi-p\,\partial_p\psi=0$, which is satisfied by $\psi=np$.

This is equivalent to the symbolic construction used by e.g.~\cite{Catchpole1997,Catchpole1998} and summarised by~\cite{Cole2010,Cole2020}, to obtain nonlinear, at least locally identifiable reparameterisations. We can also formulate the same construction from the row-space side, by equating the rows of $D_\theta\psi(\theta)$ to vectors spanning $\rowspace D_\theta\phi(\theta)$. For the resulting reparameterisation to be globally valid on the chosen parameter domain, $\tilde{\phi}$ must further be one-to-one on the entire reduced-coordinate domain $\psi(\Theta)$.

The above provides conditions to solve for a minimal image reparameterisation. In some cases we may seek an image reparameterisation that is not minimal. We can then derive an image reparameterisation satisfying the subset conditions by enforcing the orthogonality condition, $D_\theta\psi(\theta)\alpha_j(\theta)=0$, for a set of vectors $\{\alpha_j\}$ that span only a proper subspace of the full model's null space. Any solution with no additional null directions beyond the chosen subspace provides a candidate non-minimal image reparameterisation. We discuss the choice of an appropriate subspace and the utility of this condition in the context of numerical approaches below.

The downside of these symbolic approaches is that they must be solved symbolically, i.e., enforced for all $\theta$ (in both the image and minimal image cases), which requires symbolic computation and is typically only possible for simple models. We again emphasise that this symbolic approach is essentially equivalent to the approaches reviewed by~\cite{Cole2010, Cole2020}, building on e.g. \cite{Catchpole1997,Catchpole1998}, though we use slightly different terminology and notation. We next show how this can be converted to a numerical approach under certain conditions.

\subsection{Numerical reparameterisation in transformed coordinates}
We now consider when the symbolic conditions above can be replaced by a numerical calculation at a single reference point. The primary obstacle is that the row and null spaces of the local derivative are typically parameter-dependent. Thus, we want to find equivalent constraints on our reparameterisation that are independent of $\theta$, so that we can evaluate them at a single convenient reference point. This is possible when the parameter dependence of the relevant row and null spaces can be removed by working in transformed coordinates.

From above, our general reparameterisation is written $\psi(\theta)$ with $\phi(\theta) = \tilde{\phi}(\psi(\theta))$. Here, we consider a restricted class. First, define transformed coordinates
\begin{equation}
\theta^* = f(\theta)
\end{equation}
where, in the main construction used here, $f$ is a smooth, one-to-one componentwise transformation with non-zero scalar derivative on the parameter domain. Equivalently, the Jacobian of the full vector transformation is diagonal and non-singular. More generally, the same chain-rule argument can be formulated for any smooth one-to-one change of coordinates with non-singular Jacobian; we focus on componentwise choices because they have a simple interpretation and make the log-monomial case transparent.

Next, define linearly-reduced, transformed coordinates
\begin{equation}
\eta = A\theta^*
\end{equation}
where $A \in \mathbb{R}^{r\times p}$ is a constant (parameter-independent), full row rank matrix with $r \leq p$ (i.e. a `wide' matrix). Thus we have a general componentwise, invertible transformation, followed by a matrix representing a reduction in dimension in transformed coordinates.

We then write the auxiliary mapping in transformed `$f$-coordinates' as
\begin{equation}
\phi_*(\theta^*) = \phi(f^{-1}(\theta^*)),
\end{equation}
where $f^{-1}(\theta^*) = \theta$. Our analysis will use derivatives in these coordinates rather than original $\theta$ coordinates.

The key structural assumption in our approach is that we can write the transformed auxiliary mapping in terms of reduced transformed coordinates:
\begin{equation}
\phi_*(\theta^*) = \bar{\phi}(A\theta^*) = \bar{\phi}(\eta).
\label{eq:reparam-structured}
\end{equation}
Here we use $\bar{\phi}$ rather than $\tilde{\phi}$ as in the symbolic section because we may include an additional one-to-one transformation linking $\eta$ and $\psi$, as we show later, but this is not important at this stage of analysis.

As in the general symbolic case we consider the chain rule, here giving:
\begin{equation}
D_{\theta^*} \phi_*(\theta^*) = D_{\eta}\bar{\phi}(\eta)A.
\label{eq:chainrule-structured}
\end{equation}
Hence, for this representation, we have the conditions:
\begin{equation}
    \begin{aligned}
\nullspace A &\subseteq \nullspace D_{\theta^*}\phi_*(\theta^*),\\
\rowspace A &\supseteq \rowspace D_{\theta^*}\phi_*(\theta^*).
\end{aligned}
\end{equation}
This is essentially the same result as before but now tells us that there are \textit{constant} subspaces satisfying these conditions, as $A$ is a constant matrix. We will demonstrate the utility of this for the minimal image case, but we also benefit in the non-minimal case.

In particular, when $D_{\eta}\bar{\phi}(\eta)$ is injective, so that the reduced map introduces no further local redundancy, we have a locally minimal image reparameterisation. For a global minimal image reparameterisation, we also require $\bar{\phi}$ to be one-to-one on the reduced-coordinate domain $A f(\Theta)$. This same global qualification applies to the symbolic differential condition above; we assume it here in the minimal case.

We then have the key conditions determining the fundamental spaces of $A$:
\begin{equation}
    \begin{aligned}
\nullspace A &= \nullspace D_{\theta^*}\phi_*(\theta^*),\\
\rowspace A &= \rowspace D_{\theta^*}\phi_*(\theta^*).
\end{aligned}
\label{eq:key-condition}
\end{equation}
The importance of isolating the constant $A$ matrix is that we can use the transformed auxiliary-map Jacobian at any convenient point to determine the row and null spaces of $A$, as long as $D_{\eta}\bar{\phi}$ is injective there, and hence choose a representative $A$ up to basis choice. This reduces what would be a symbolic differential calculation to an algebraic computation. If the model is structurally non-identifiable for the chosen auxiliary mapping, the point used will generally be, e.g., any representative maximum likelihood estimate, and the Jacobian will be singular. This is expected and causes no issues in general. In the running example, with $f=\log$, the transformed Jacobian $D_{\theta^*}\phi_*(\theta^*)$ has both rows equal to $np\,(1,1)$, so its row and null spaces are the same at every point, and $A=(1,1)$ gives $\eta=\log n+\log p=\log(np)$.

The exact single-point calculation here comes from the representation $\phi_*(\theta^*)=\bar{\phi}(A\theta^*)$ with $A$ constant. If it is known \textit{a priori} that the full local null space is invariant (constant) given this representation, then the local Jacobian already provides the required row and null spaces. In general, however, we may not know whether the full local null space is invariant, or whether only a proper subspace is invariant. We therefore need a way to extract the invariant component of the local null space and to check whether its dimension matches that of the full null space. This gives the minimal case when the dimensions agree, and a non-minimal image reparameterisation when the invariant component is smaller. We later use a first-order invariance calculation to extract candidate invariant directions for this purpose.

While the above condition does not uniquely determine $A$, as any basis for the relevant null space and complementary row space can be used, it establishes the existence of a matrix capturing the non-identifiable directions and the retained, potentially identifiable parameter combinations. In the log-transformed case, rows of $A$ correspond to exponent vectors for monomial parameter combinations. Such exponents are traditionally restricted to integers in dimensional analysis, but here we allow them to be arbitrary real numbers and do not require the combinations to be dimensionless \textit{a priori}. This relaxation facilitates numerical computation and provides more flexibility in discovering practically useful parameter combinations. However, we also consider methods for enforcing integer constraints and providing more interpretable parameter combinations.

In the case of an image reparameterisation, the core reparameterisation
idea can also be applied sequentially: the reduced coordinates obtained
from one image reparameterisation can be treated as the starting
parameterisation for a further reparameterisation, using a new $f$ and
$A$. We illustrate this briefly in the present work as a way to understand non-minimal images, but leave full development of sequential applications to future work. Next we address another source of non-uniqueness in the reparameterisation, beyond the choice of $A$.

\subsection{Coordinate choice on the invariant image}
There are multiple sources of non-uniqueness in the reparameterisation, even for a minimal image reparameterisation of the form above. We have both non-uniqueness in the reparameterisation given a fixed $A$ representation and non-uniqueness in the representation of $A$. First, we consider the former.

Given a choice of $\eta$, the same argument as above goes through if we use a reparameterisation of the form
\begin{equation}
\psi(\theta) = g(Af(\theta)) = g(A\theta^*)= g(\eta),
\end{equation}
where $g$ is a smooth, one-to-one componentwise transformation with non-zero scalar derivative on the reduced-coordinate domain. That is, we can include a further one-to-one reparameterisation of the linearly reduced, transformed coordinates without difficulty. This means we have the relationship
\begin{equation}
\tilde{\phi} \circ g = \bar{\phi}.
\end{equation}
Though unneeded for the general theory, this is natural in the primary case of interest we consider. This case of interest takes the form:
\begin{equation}
\psi(\theta) = g \circ A \circ f = \exp \circ A \circ \log (\theta) = \exp(A \log \theta),
\end{equation}
where $g = \exp$ and $f = \log$ are applied as componentwise functions of vector inputs, and in fact $g = f^{-1}$ so $\psi = f^{-1} \circ A \circ f$. In the running example, this recovers $\psi=\exp(\eta)=np$, and the null direction $(1,-1)$ gives the complementary coordinate $\exp(\log n-\log p)=n/p$.

Such transformations generate vectors of monomial combinations of the entries of an input vector, i.e., each entry of $\psi(\theta)$ is a monomial of the form
\begin{equation}
\psi_j(\theta) = \theta_1^{A_{j1}}\theta_2^{A_{j2}}\cdot\dots\theta_p^{A_{jp}}
\end{equation}
where $A_{ji}$ are the entries in the $j$th row of $A$. These powers are integers for strict monomials, though we allow non-integer values in general. Here we assume that the sign of each entry of $\theta$ is either known or can be estimated (via the maximum likelihood estimate, for example). Hence, we can apply the log function to the absolute value of each entry and then reintroduce the sign in the auxiliary mapping if needed.

Log parameter transformations are also commonly used in the sloppiness literature~\cite{Brown2004,Monsalve2022,Vollert2023} and this also partly inspired our present approach, though our approach explicitly treats the log transformation as part of general structural reparameterisation transformations.

\begin{figure}[htbp]
    \centering
    \includegraphics[width=0.6\linewidth]{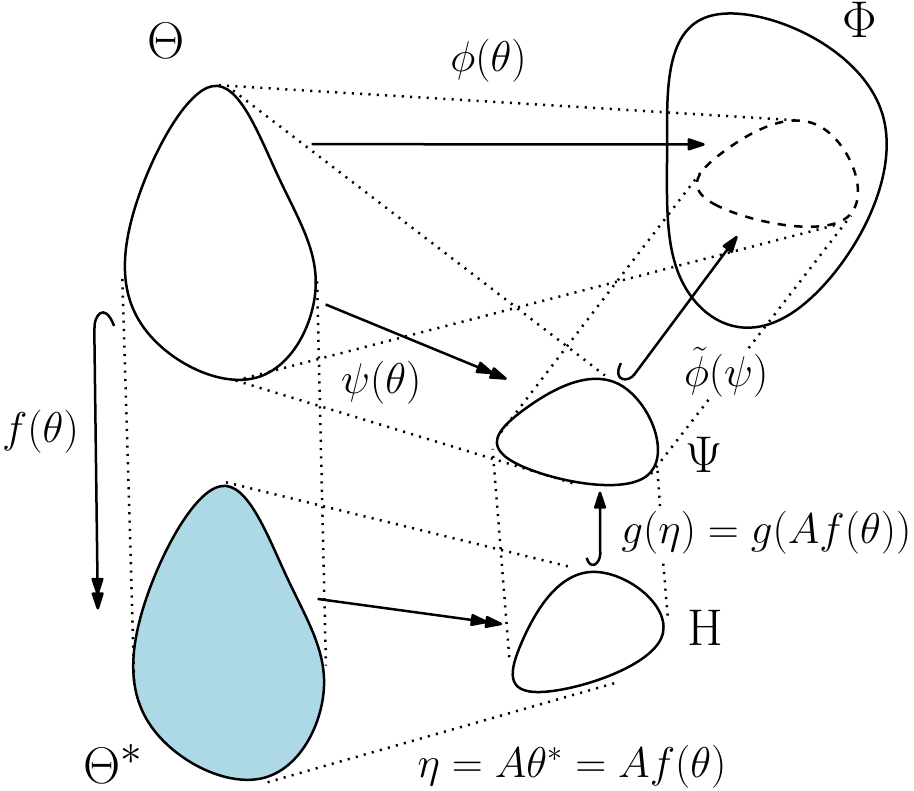}
    \caption{The upper route shows the basic image decomposition $\phi(\theta)=\tilde{\phi}(\psi(\theta))$, while the lower route shows the transformed-coordinate construction used in IIR. An initial componentwise change of variables $f(\theta)$, typically $\log$, is followed by a constant linear reduction $A$ and an optional componentwise reparameterisation $g$ of the reduced coordinates $\eta=A f(\theta)$, typically $g=f^{-1}=\exp$. Under suitable assumptions, the chain rule along the path from $\Theta^*$ to $\Phi$ relates the row and null spaces of the auxiliary-map Jacobian to those of the constant matrix $A$, with equality in the minimal case. Double arrows indicate onto maps and hooked arrows indicate one-to-one maps, following~\cite{Lawvere2003}.}
    \label{fig:epi-linear}
\end{figure}

The basic image decomposition and the transformed-coordinate route used for IIR are illustrated in Figure~\ref{fig:epi-linear}.

\subsection{First-order invariance calculation for invariant null-space components}
The representation $\phi_*(\theta^*)=\bar{\phi}(A\theta^*)$, with $A$ constant, guarantees that the null directions associated with the reduced coordinates are constant; in the minimal case, these coincide with the full local null space. In general, however, the full local null space of $D_{\theta^*}\phi_*(\hat{\theta}^*)$ need not be invariant: in non-minimal image cases only a proper subspace of it may remain constant. We therefore use a first-order invariance calculation to extract candidate invariant null directions. Here we summarise the key results; full derivation is given in the Supplementary Material.

\begin{samepage}
Let $\hat{\alpha}$ be a null-space vector of $D_{\theta^*}\phi_*(\hat{\theta}^*)$. A necessary first-order condition for $\hat{\alpha}$ to remain a constant null direction under small perturbations is
\begin{equation}
H_i(\hat{\theta}^*)\hat{\alpha}=0,\qquad i=1,\dots,p,
\label{eq:invariance-condition}
\end{equation}
where $H_i(\hat{\theta}^*)=D_{\theta_i^*}D_{\theta^*}\phi_*(\hat{\theta}^*)\in\mathbb{R}^{q\times p}$ is the derivative of the Jacobian with respect to the $i$th transformed parameter.

\end{samepage}
This condition is obtained by expanding the Jacobian
\begin{equation}
D_{\theta^*}\phi_*(\hat{\theta}^*+\delta)
=
D_{\theta^*}\phi_*(\hat{\theta}^*)+
\sum_i H_i(\hat{\theta}^*)\delta_i+O(\|\delta\|^2)
\end{equation}
and requiring the null condition to hold to first order for all small perturbations $\delta$. Thus failure of \eqref{eq:invariance-condition} rules out a constant null direction at first order. Passing the condition at a single point does not, by itself, show that a direction is exactly constant, even when a constant-$A$ representation holds. Directions that pass are candidates, valid to first order, which can be further checked, for example, by evaluating the Jacobian on these fixed directions at points away from the reference point, or by additional analytical arguments (as in the extended example below).

To compute candidate invariant directions in the local null space, let $V_0$ denote the right singular vectors spanning the local null space, and write any local null-space vector as $\hat{\alpha}=V_0c$. Combining \eqref{eq:invariance-condition} for all parameters gives
\begin{equation}
\begin{bmatrix}
H_1(\hat{\theta}^*)V_0\\
\vdots\\
H_p(\hat{\theta}^*)V_0
\end{bmatrix}C_0=0,
\end{equation}
where the columns of $C_0$ are coefficient vectors for those local null
directions that satisfy the first-order invariance condition. Computationally, the invariance calculation uses derivatives of the transformed auxiliary-map Jacobian, but only through the products $H_i(\hat{\theta}^*)V_0$ appearing in the stacked matrix. In the implementation these products are obtained by differentiating the Jacobian--null-basis product
$\theta^* \mapsto D_{\theta^*}\phi_*(\theta^*)V_0$, with $V_0$ held fixed,
using automatic differentiation. This is purely for efficiency, as it avoids constructing the full Hessian tensor. The same products could also be obtained by computing all Hessian entries explicitly, or approximately by finite differences of the Jacobian if automatic differentiation is not available, for example when the model is evaluated by external software, though step sizes and solver tolerances then need care.

A default choice for $C_0$ is given by the right singular vectors associated with zero singular values of this stacked matrix; denoting these vectors by $V_{M0}$, a basis for the candidate invariant null directions is $N=V_0V_{M0}$. The remaining local null directions fail this first-order invariance test; they are given by the columns of $V_0V_{Mr}$, where the columns of $V_{Mr}$ are the right singular vectors of the stacked matrix associated with non-zero singular values. We therefore include them in the canonical complement to the candidate invariant null space, concatenating these column bases as
\begin{equation}
N_\perp=\begin{bmatrix} V_r & V_0V_{Mr}\end{bmatrix},
\end{equation}
where $V_r$ spans the local row space from the original Jacobian SVD. If $N$ spans the full local null space, the result is consistent with a minimal image reparameterisation. This can be confirmed under the additional assumptions given above, of a constant-$A$ representation whose reduced map has an injective derivative. If $N$ is lower-dimensional, part of the local null space is not constant, so no minimal reparameterisation of this form exists. The complement $N_\perp$ is then used to construct the default reduction matrix $A$; when the directions in a lower-dimensional $N$ are exactly invariant (not just to first order), we get an exact but non-minimal image reparameterisation. Whether or not these directions are invariant, keeping all coordinates gives an exact, invertible change of coordinates (see Section~\ref{sec:practical-reparam}).

\subsection{Overall algorithm}
The overall approach is summarised in the following algorithm.
\begin{algorithm}[H]
\caption{Invariant-image computation and canonical reparameterisation}
\label{alg:reparam}
\begin{algorithmic}[1]
\Input Auxiliary mapping $\phi(\theta)$, reference parameter value $\hat{\theta}$, componentwise transformations $f$ (e.g., $f=\log$), $g$. Numerical tolerances for determining zero singular values.
\Output Candidate invariant null basis $N$, canonical orthonormal complementary basis $N_\perp$, and default reparameterisation matrix $A_{\mathrm{SVD}} = N_\perp^T$
\State Define $\theta^* = f(\theta)$
\State Compute transformed reference point $\hat{\theta}^* = f(\hat{\theta})$
\State Compute the Jacobian matrix $D_{\theta^*}\phi_*(\hat{\theta}^*)$ numerically
\State Compute the SVD $D_{\theta^*}\phi_*(\hat{\theta}^*) = U\Sigma V^T$
\State Partition $V=\begin{bmatrix}V_r & V_0\end{bmatrix}$ based on the non-zero and zero singular values
\State Compute the products $H_i(\hat{\theta}^*)V_0$ needed in \eqref{eq:invariance-condition} numerically, and form the vertically stacked matrix $M_{\text{stack}} = [H_1(\hat{\theta}^*) V_0; H_2(\hat{\theta}^*) V_0; \dots; H_p(\hat{\theta}^*) V_0]$
\State Compute the SVD of $M_{\text{stack}}$ and partition the right singular vectors of $M_{\text{stack}}$ into $V_M=\begin{bmatrix}V_{Mr} & V_{M0}\end{bmatrix}$ based on non-zero and zero singular values
\State Construct bases $N = V_0V_{M0}$ for the candidate invariant null directions and $N_{\perp} = \begin{bmatrix} V_r & V_0V_{Mr}\end{bmatrix}$ for its complement
\State Set the default reparameterisation matrix $A_{\mathrm{SVD}} = N_{\perp}^T$
\State Return $N$, $N_{\perp}$, and $A_{\mathrm{SVD}}$ for the default reduced transformed reparameterisation $\eta_{\mathrm{SVD}} = A_{\mathrm{SVD}} \circ f$, and possibly full image reparameterisation $\psi_{\mathrm{SVD}} = g \circ A_{\mathrm{SVD}} \circ f$.
\end{algorithmic}
\end{algorithm}
Algorithm~\ref{alg:reparam} computes the candidate invariant null basis $N$ and its canonical orthonormal complement $N_\perp$. When the directions in $N$ are exactly invariant, the associated SVD reparameterisations $\eta_{\mathrm{SVD}}(\theta)=A_{\mathrm{SVD}}f(\theta)$ and $\psi_{\mathrm{SVD}}(\theta)=g(A_{\mathrm{SVD}}f(\theta))$ are image reparameterisations, minimal or non-minimal depending on whether $N$ spans the full local null space or only a subspace of it, as described above. The reparameterised model can then be used for parameter estimation and uncertainty quantification. Other reparameterisation matrices may be chosen to obtain more interpretable coordinates while spanning the same image (see below).

When $V_{Mr}$ is empty, equivalently when $\rank(N)= \rank(V_0)$, every local null direction passes the invariance test; otherwise $N_\perp$ retains local null directions failing the invariance test, and no minimal reparameterisation of this form exists.

\subsection{Practical reparameterisation}\label{sec:practical-reparam}
In structurally identifiable but practically weakly informed settings, the Jacobian may be full rank but have singular values of very different magnitudes. In this case IIR does not give an exact dimension reduction. Instead, the SVD basis and any sparse basis chosen within the same transformed parameter space provide image coordinates ordered, or interpreted, using local information content. Weak coordinates should therefore be interpreted as weakly informed image coordinates, not as exactly non-identifiable null coordinates. Even after an exact reduction, retained coordinates with small but non-zero singular values can be poorly estimated from finite, noisy data, so for practical purposes the split into strongly and weakly informed coordinates may be more useful.

For computation it can still be useful, in both structurally and practically non-identifiable cases, to keep a full square reparameterisation, combining coordinates associated with $N_\perp$ and, when present, coordinates associated with $N$. This is an invertible change of coordinates, so it represents the model exactly whether or not the directions in $N$ are invariant. It also preserves compatibility with existing simulation code while making the locally identifiable, non-identifiable, or weakly informed directions explicit. Fixing the coordinates associated with $N$ then gives an exact reduction when the observable behaviour does not depend on these coordinates over the allowed parameter domain, and the fixed values remain allowed by the transformed parameter constraints across the retained range. The inverse reparameterisation maps the retained coordinates and these fixed values back to the original parameters, so the reduced model can be simulated with the original code. Fixing weakly informed coordinates instead gives an approximate reduction, which can be interpreted as a model driven mainly by the strongly informed coordinates. Its accuracy depends on the quantities of interest and needs to be assessed case by case.

\subsection{Practical vs structural identifiability and observation operators}
For temporal or spatial models, identifiability also depends on the observation operator. If $\phi_{\mathrm{fine}}$ denotes a fine-grid model solution and
\begin{equation}
\phi_{\mathrm{obs}}=B_{\mathrm{obs}}\phi_{\mathrm{fine}},
\end{equation}
then parameters identifiable from the fine-grid solution may become weakly informed or non-identifiable after projection to observed quantities or a coarser observation grid. Conversely, augmenting the observation map can remove null directions. The same calculation applies to a prediction or candidate measurement $q$, written as a function of the transformed parameters $\theta^*$. To first order, if $q$ is unchanged along every local null direction of the current observation map, then it depends only on the identifiable combinations. Otherwise, adding $q$ to the observation map removes null directions; as shown in the Supplementary Material, the number removed is the rank of the derivative of $q$ along them. We apply this to the unobserved repressilator proteins in Section~\ref{sec:repressilator-results}. Here we take the auxiliary mapping used for IIR to match the observable quantities of interest but usually use the fine-grid solution to approximate the structural identifiability setting. Further details on the relation between auxiliary mappings, observed information, and observation operators are given in the Supplementary Material.

\subsection{Basis choice: SVD and sparse monomial bases}
Algorithm~\ref{alg:reparam} computes a default reduction matrix $A_{\mathrm{SVD}}=N_\perp^T$. The invariant image itself is basis-independent: any full-row-rank matrix $A$ with
\begin{equation}
\rowspace A=\rowspace A_{\mathrm{SVD}}
\end{equation}
gives an equivalent image representation, differing only in the coordinates used within the image.

The default SVD construction is useful because it is orthonormal and ordered by local singular values. In high-dimensional problems, however, the corresponding parameter combinations may be dense and difficult to interpret mechanistically. We therefore also consider sparse monomial bases, obtained by a heuristic search over a bounded dictionary of integer exponent vectors in log-parameter space and selected to span the same target subspace. On the identifiable side, initial sparse candidates are ordered using local information from the Jacobian; on the invariant-null side, where local information scores are zero, candidates are ordered lexicographically by measures of simplicity, starting with support size (number of non-zero terms). Full details are given in the Supplementary Material. Sparse bases are not unique and need not preserve the SVD ordering, but can give simpler and more interpretable parameter combinations while spanning the same invariant image or invariant-null subspace. This simple search heuristic is related to methods appearing in the sparse-basis and sparse-vector-in-a-subspace literature~\cite{GilbertHeath1987,QuSunWright2016}. More sophisticated basis-selection algorithms are possible, but the heuristic used here sufficed to find interpretable bases for the examples in this article and additional examples in the repository.

Bases can also be chosen for estimation, for example so that the coordinates are closer to orthogonal in the local information matrix, which depends on the observation and noise model as well as on the auxiliary map. In particular, interest parameters can be kept for their scientific meaning while the nuisance coordinates are chosen to be orthogonal, or nearly so, to the interest parameters, as in the parameter orthogonality approach of Cox and Reid~\cite{Cox1987} and Aitkin and Stasinopoulos~\cite{Aitkin1989}. For example, when the auxiliary map is the mean of observations with independent Gaussian errors of equal variance, the Fisher information is proportional to $J^TJ$, where $J=D_{\theta^*}\phi_*$, so the SVD coordinates are orthogonal in the Fisher information at the reference point. In our examples, we choose sparse bases mainly for interpretation.

\subsection{Profile-wise analysis}
We quantify uncertainty using our PWA framework~\cite{Simpson2023}, which connects identifiability, parameter estimation, and predictive uncertainty via likelihood-based confidence sets. This approach uses both the joint likelihood for all parameters and profile likelihoods, where the likelihood is evaluated as a function of a target parameter while maximising over the other `nuisance' parameters for each value of the target parameter~\cite{Kreutz2013,Raue2014,Cox2006,Pawitan2001}. We implement our methods numerically in Julia, with full details, including the packages used, available in the \href{https://github.com/omaclaren/reparam/}{repository}.

As the likelihood function is invariant to one-to-one reparameterisation~\cite{Pace1997, Pawitan2001, Cox2006}, reparameterisation of the full likelihood (including both identifiable and non-identifiable parameters) preserves all information in the original likelihood, unlike quadratic approximations based on the Fisher information. Reparameterisation simply presents the information in the likelihood in a clearer form, while profile likelihoods reveal identifiability of individual parameters, or parameter combinations, separately from nuisance parameters~\cite{Aitkin1989}.

\section{Example models}
\label{sec:Models}
Here we consider two examples. The first is a simple parameterised normal model, used as an elementary case for which the main steps of the method can be followed analytically as well as numerically. We include variations on this model including structurally non-identifiable and practically non-identifiable cases, and a case where the standard approach gives a non-minimal image reparameterisation. In this latter case we show how a sequential application of the approach can obtain the minimal reparameterisation. The second example is the repressilator, a nonlinear ODE model of a synthetic gene regulatory network~\cite{Elowitz2000,Eisenberg2014}, used to show that the same approach can detect the non-identifiability of the original parameters and recover a reduced parameterisation with identifiable image coordinates. This example also shows that the simple search heuristic for sparse monomial bases can provide more interpretable parameter combinations than the default SVD basis. Additional implementation examples, including Michaelis--Menten kinetics and a heterogeneous-flow/transport model, are available in the accompanying \href{https://github.com/omaclaren/reparam}{code repository} and summarised briefly in the Supplementary Material.

\subsection{Parameterised normal approximations}
Our first example, in its simplest form, involves estimating the number of trials $n$ and success probability $p$ in a continuous approximation to a binomial model. Near the Poisson limit (large $n$, small $p$), maximum likelihood estimates become unstable~\cite{Olkin1981}, providing a case study in practical non-identifiability. Although relatively simple, this model (in its discrete form) was recently described~\cite{Murph2024} as `yet to receive a satisfactory solution using any [statistical] philosophy whatsoever'. We share the general philosophy, discussed in e.g.~\cite{Davison2006}, that embedding a discrete statistical model in a continuous one can be a very useful approach to inference, enabling more sophisticated inference methods relying on derivatives, and here assume our variables (including $n$) can be approximated by continuous ones and that sufficient differentiability holds. We consider both the non-limiting and the limiting Poisson cases explicitly. We then extend the limiting Poisson case to a scenario that represents a sum of Poisson limit models, with additional non-identifiability in the form of a sum of monomial parameters.

The model for a single (`$n$-trial') experiment is:
\begin{equation}
Y \sim \mathcal{N}(np, np(1-p)),
\label{eq:normal-bin}
\end{equation}
with Poisson limit case:
\begin{equation}
Y \sim \mathcal{N}(np, np).
\label{eq:normal-poi}
\end{equation}
The auxiliary mapping connects the underlying $(n,p)$ `mechanistic' parameters to normal distribution parameters:
\begin{equation}
\phi: \begin{bmatrix}
    n\\
    p
\end{bmatrix}
\mapsto \begin{bmatrix}
    np\\
    np(1-p)
\end{bmatrix}
= \begin{bmatrix}
    \mu\\
    \sigma^2
\end{bmatrix}.
\end{equation}
Although this is not really a `mechanistic' model in the usual sense, one could view an underlying binomial model as a lower-level mechanism that `generates' the higher-level (approximately) normally distributed observations. The problem is to determine both $n$ and $p$ given $k$ observations from the single experiment model given by \eqref{eq:normal-bin} or \eqref{eq:normal-poi}.

In the case of a sum of two independent Poisson limit models, we have:
\begin{equation}
Y \sim \mathcal{N}(n_1p_1 + n_2p_2, n_1p_1 + n_2p_2),
\label{eq:normal-poi-sum}
\end{equation}
with auxiliary mapping:
\begin{equation}
\phi: \begin{bmatrix}
    n_1\\
    p_1\\
    n_2\\
    p_2
\end{bmatrix}
\mapsto \begin{bmatrix}
    n_1p_1 + n_2p_2\\
    n_1p_1 + n_2p_2
\end{bmatrix}
= \begin{bmatrix}
    \mu\\
    \sigma^2
\end{bmatrix}.
\end{equation}
This model has a different class of structural non-identifiability: only the sum $n_1p_1+n_2p_2$ is identifiable, not the individual products or their component parameters.

\subsection{Repressilator model}
Our second example is the repressilator, a synthetic three-gene regulatory network introduced by Elowitz and Leibler~\cite{Elowitz2000} and later used as an example for identifiability analysis by Eisenberg and Hayashi~\cite{Eisenberg2014}. We use the same model structure as Eisenberg and Hayashi, but fix the Hill coefficient at $n=2.5$ and use a different synthetic parameter set for the present illustrative study. Eisenberg and Hayashi considered the corresponding version with the Hill coefficient $n$ treated as a fitted parameter, noting that the fitted Hill exponent makes the usual differential-algebra approach inapplicable. This matters because differential algebra and related elimination methods are an important class of symbolic structural-identifiability methods, as implemented for example in DAISY~\cite{Bellu2007} and Gröbner-basis approaches to finding identifiable combinations~\cite{Meshkat2009}; see also Chis et al.~\cite{Chis2011} for a comparison of methods. These methods are typically formulated for ODE models with right-hand sides and outputs that are polynomial or rational functions of the states and parameters. Even with $n$ fixed at a non-integer value, as here, the Hill terms are not polynomial or rational functions of the state variables and parameters in the form required by these standard approaches, without additional transformations. This limitation is specific to that class of symbolic elimination methods: the symbolic invariance condition motivating IIR can be stated for any sufficiently smooth auxiliary map $\phi$, provided that the map and its derivatives are available. The numerical IIR calculation uses derivatives of a computable auxiliary map, computed here by automatic differentiation, so it can still be applied when tractable symbolic expressions are unavailable or when the model is not of rational form. We also include a supplementary IIR check in the repository, treating $n$ as unknown, and show that the same invariant-image calculation can still be applied in this case. For simplicity, however, we compute the full profile-wise analysis only for the fixed-$n$ case. Full parameter values, bounds, initial conditions, and observation settings are given in the Supplementary Material.

The noise-free latent state vector is given by
\begin{equation}
X = [m_1,m_2,m_3,p_1,p_2,p_3]^T,
\end{equation}
where $m_i$ and $p_i$ denote mRNA and protein concentrations. As in~\cite{Eisenberg2014}, we assume that only the mRNA trajectories are observed.

The 18 model parameters are the basal transcription rates $\alpha_{0i}$, regulated transcription rates $\alpha_i$, translation rates $\beta_i$, inhibition constants $K_i$, mRNA degradation rates $k_{\mathrm{degm},i}$, and protein degradation rates $k_{\mathrm{degp},i}$, for $i=1,2,3$. The model equations are

\begin{equation}
\begin{aligned}
\frac{dm_1}{dt} &= \alpha_{01} + \frac{\alpha_1}{1 + (p_3/K_3)^n} - k_{\mathrm{degm},1} m_1,\\
\frac{dm_2}{dt} &= \alpha_{02} + \frac{\alpha_2}{1 + (p_1/K_1)^n} - k_{\mathrm{degm},2} m_2,\\
\frac{dm_3}{dt} &= \alpha_{03} + \frac{\alpha_3}{1 + (p_2/K_2)^n} - k_{\mathrm{degm},3} m_3,\\
\frac{dp_1}{dt} &= \beta_1 m_1 - k_{\mathrm{degp},1} p_1,\\
\frac{dp_2}{dt} &= \beta_2 m_2 - k_{\mathrm{degp},2} p_2,\\
\frac{dp_3}{dt} &= \beta_3 m_3 - k_{\mathrm{degp},3} p_3.
\end{aligned}
\label{eq:repressilator}
\end{equation}
For the invariant-image calculation, the auxiliary mapping takes $\theta$ to the stacked three-mRNA solution on a fine time grid. For the likelihood-based analysis, we assume the mRNA trajectories are observed at a smaller set of time points, with additive Gaussian observation error:
\begin{equation}
y_{\mathrm{obs}} \sim \mathcal{N}(B_{\mathrm{obs}}m_{\mathrm{fine}}(\theta), \sigma^2 I),
\end{equation}
where $m_{\mathrm{fine}}(\theta)$ denotes the stacked fine-grid mRNA solution and $B_{\mathrm{obs}}$ maps this to the observation times.

\section{Results and Discussion}
Here we present the results of our analysis of the two classes of model described in Section~\ref{sec:Models}. We use the parameterised normal model to illustrate the main ideas in a simple setting, including the distinction between symbolic and numerical approaches to reparameterisation, the distinction between non-minimal and minimal image reparameterisation, and the application of the approach to practically non-identifiable models. We then show how the same approach can be applied to a larger nonlinear mechanistic model, the repressilator. For this last example, we show how parameter non-identifiability and predictive uncertainty are related using the PWA approach.

\subsection{Parameterised Normal Models}
These examples are sufficiently simple that we can illustrate the core ideas analytically, and so we first outline some analytical results. We then present numerical results, produced without assuming these analytical results were available, using generic model-agnostic code.

We focus our analytical illustration on the Poisson limit model for simplicity, defined by \eqref{eq:normal-poi} and, for the extended parameterisation, \eqref{eq:normal-poi-sum}. We show both symbolic and invariant image approaches to reparameterisation, including the non-minimal reduction that arises in the extended case.

In addition to analytical results, we present numerical, likelihood-based analysis of both the Poisson-limit and non-limit versions of the basic parameterised normal model. We consider the likelihood functions in the original parameterisation and in the reparameterised coordinates.

\subsubsection{Analytical summary: Poisson-limit models}
Here, we summarise the analytical results for both the simple and extended Poisson-limit models. The full symbolic and first-order invariance calculations are given in the Supplementary Material.

First, for the auxiliary map of the basic Poisson-limit model,
\begin{equation}
\phi(n,p)=
\begin{bmatrix}
np\\
np
\end{bmatrix},
\end{equation}
the Jacobian in original coordinates is
\begin{equation}
D\phi(n,p)=
\begin{bmatrix}
p & n\\
p & n
\end{bmatrix}.
\end{equation}
This has a null-space vector $\alpha(n,p) = (n,-p)^T$, which depends on the parameter value. A numerical null-space calculation at one reference point therefore gives a vector that only matches the symbolic vector at the evaluation point but not elsewhere in parameter space. Following the above theory, the corresponding reparameterisation condition can be written as
\begin{equation}
D\psi \alpha = \frac{\partial \psi}{\partial n}n
+\frac{\partial \psi}{\partial p}(-p)
=0,
\end{equation}
with (non-unique) solution $\psi(n,p)=np$. This recovers the expected identifiable combination, but requires a global symbolic calculation.

Now, for the IIR analysis, we take log coordinates, $n^*=\log n$ and $p^*=\log p$, giving
\begin{equation}
\phi_*(n^*,p^*)=
\begin{bmatrix}
\exp(n^*+p^*)\\
\exp(n^*+p^*)
\end{bmatrix},
\qquad
D\phi_*(n^*,p^*)=
\exp(n^*+p^*)
\begin{bmatrix}
1&1\\
1&1
\end{bmatrix}.
\end{equation}
The row and null spaces are now constant, with unnormalised basis vectors $(1,1)$ and $(1,-1)$. Thus, up to normalisation, the constant row vector gives
\begin{equation}
A=\begin{bmatrix}1&1\end{bmatrix},
\qquad
\eta=A
\begin{bmatrix}
n^*\\
p^*
\end{bmatrix}
=n^*+p^*=\log(np).
\end{equation}
With the additional componentwise transformation $g=\exp$, this gives the image coordinate
\begin{equation}
\psi(n,p)=g(\eta)=np.
\end{equation}
The complementary constant null direction gives the invariant-null coordinate
\begin{equation}
\lambda(n,p)
=
\exp\!\left(
\begin{bmatrix}
1&-1
\end{bmatrix}
\begin{bmatrix}
\log n\\
\log p
\end{bmatrix}
\right)
=
\frac{n}{p}.
\end{equation}
Thus the same reparameterisation information obtained symbolically in the original coordinates is recovered from a single transformed-coordinate Jacobian calculation.

For the extended, two-component Poisson-limit model, where the auxiliary map depends only on $n_1p_1+n_2p_2$, the minimal symbolic parameterisation is in terms of the single variable $n_1p_1+n_2p_2$. This can be found by solving the symbolic reparameterisation condition. In contrast, a first log-monomial IIR application gives the exact but non-minimal image parameterisation in terms of
\begin{equation}
(n_1p_1,n_2p_2).
\end{equation}
While not minimal, this is still useful: it shows that the model depends on the original four parameters only through two monomial coordinates. Furthermore, the invariance test and rank information indicate clearly that further reduction is possible: only a two-dimensional subspace of the three-dimensional local null space passes the invariance test. An additional argument using the Hessian slices given in the Supplementary Material confirms that this candidate subspace is exactly invariant. This motivates a second application of IIR using this as the starting parameterisation.

Taking $f$ and $g$ as identity functions and using $(n_1p_1,n_2p_2)$ as the starting parameterisation, a second application of IIR recovers the minimal additive coordinate $n_1p_1+n_2p_2$, with complementary coordinate $n_1p_1-n_2p_2$, and indicates no further reduction is possible. This is the expected result and illustrates how invariant-image coordinates can be used as an intermediate space for further reductions. Systematic searches over such transformations are left for future work.

\subsubsection{Numerical results}
\begin{figure}[!htbp]
    \centering
    \begin{subfigure}[t]{0.32\linewidth}
        \includegraphics[width=\linewidth]{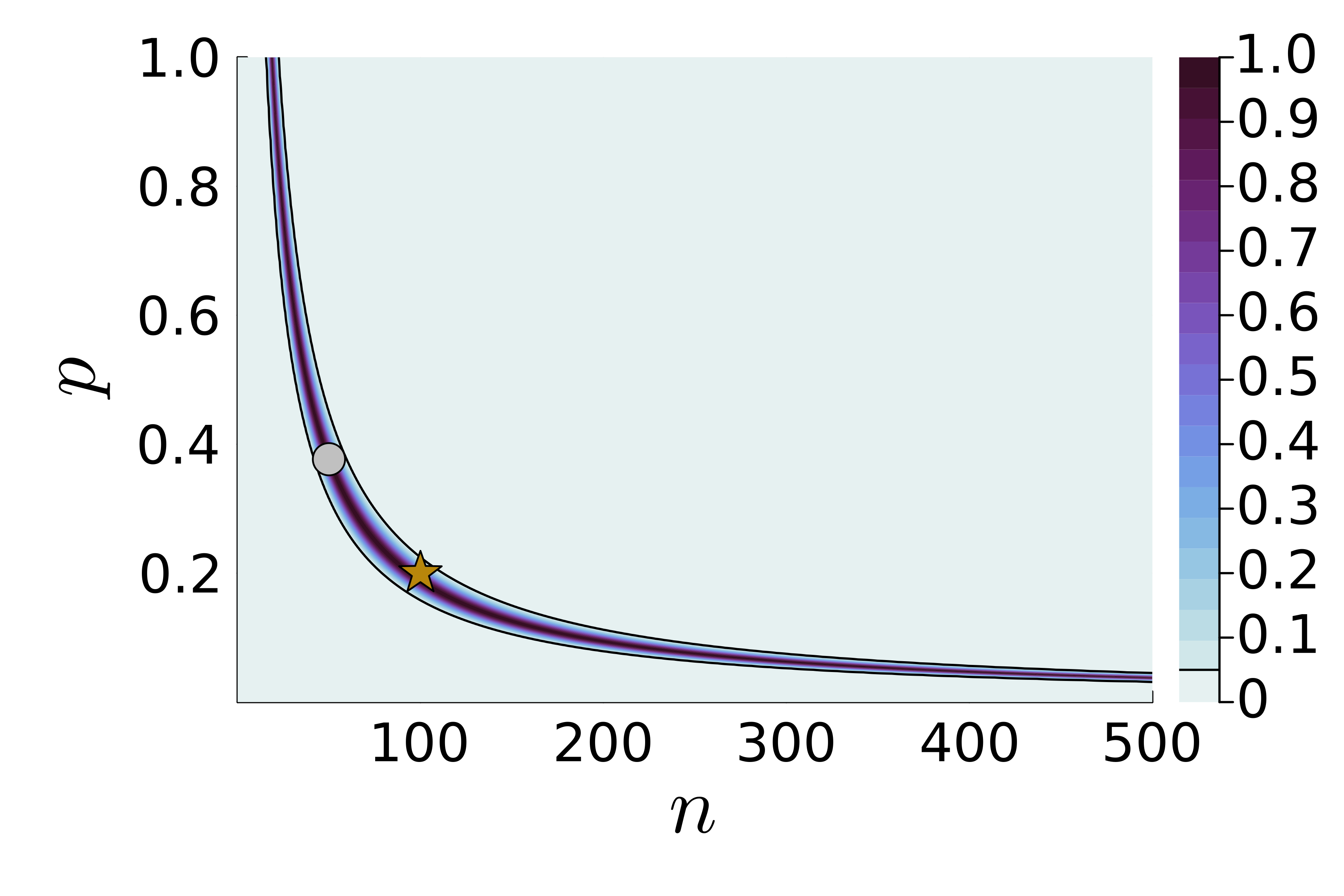}
    \end{subfigure}
    \hfill
    \begin{subfigure}[t]{0.32\linewidth}
        \includegraphics[width=\linewidth]{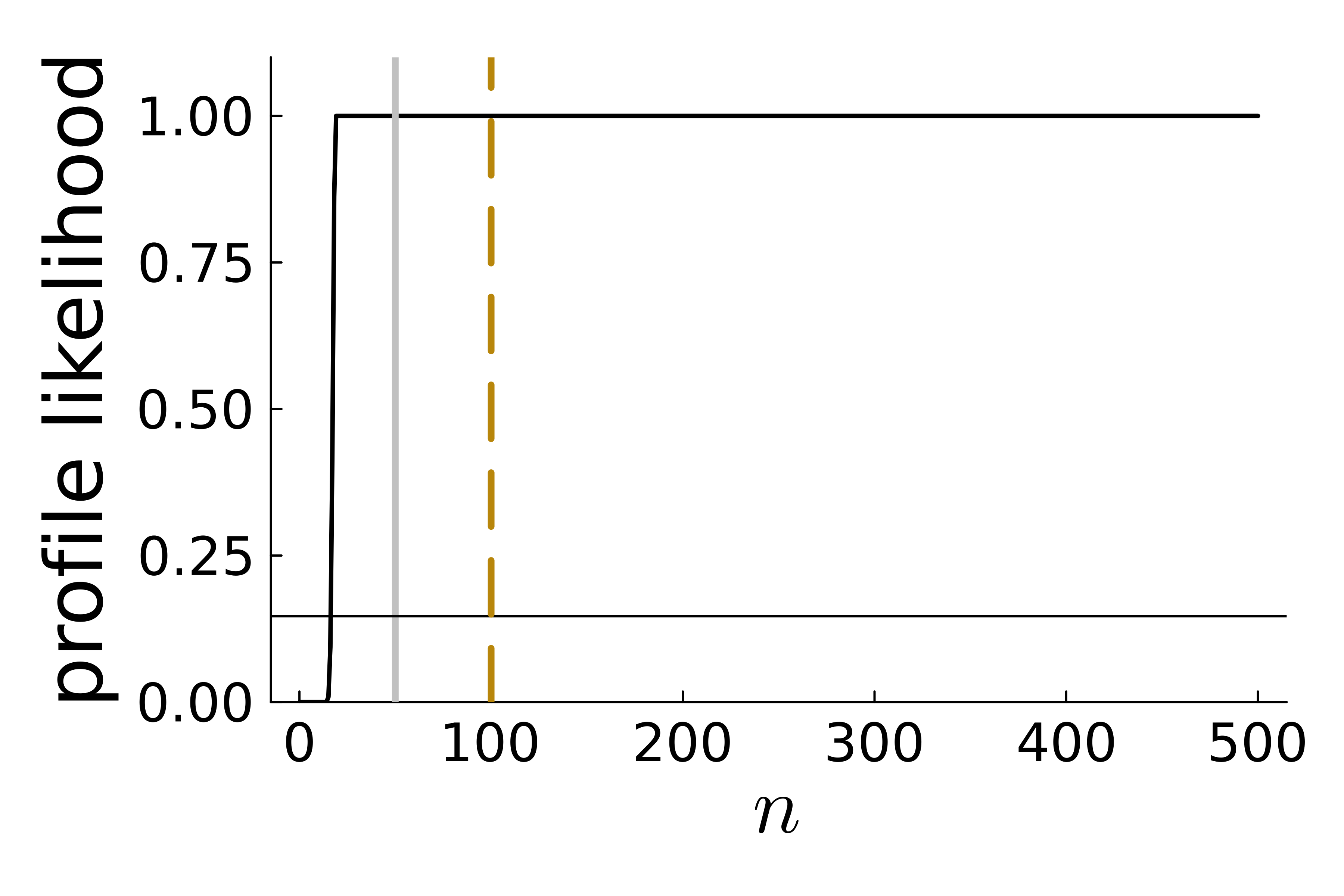}
    \end{subfigure}
    \hfill
    \begin{subfigure}[t]{0.32\linewidth}
        \includegraphics[width=\linewidth]{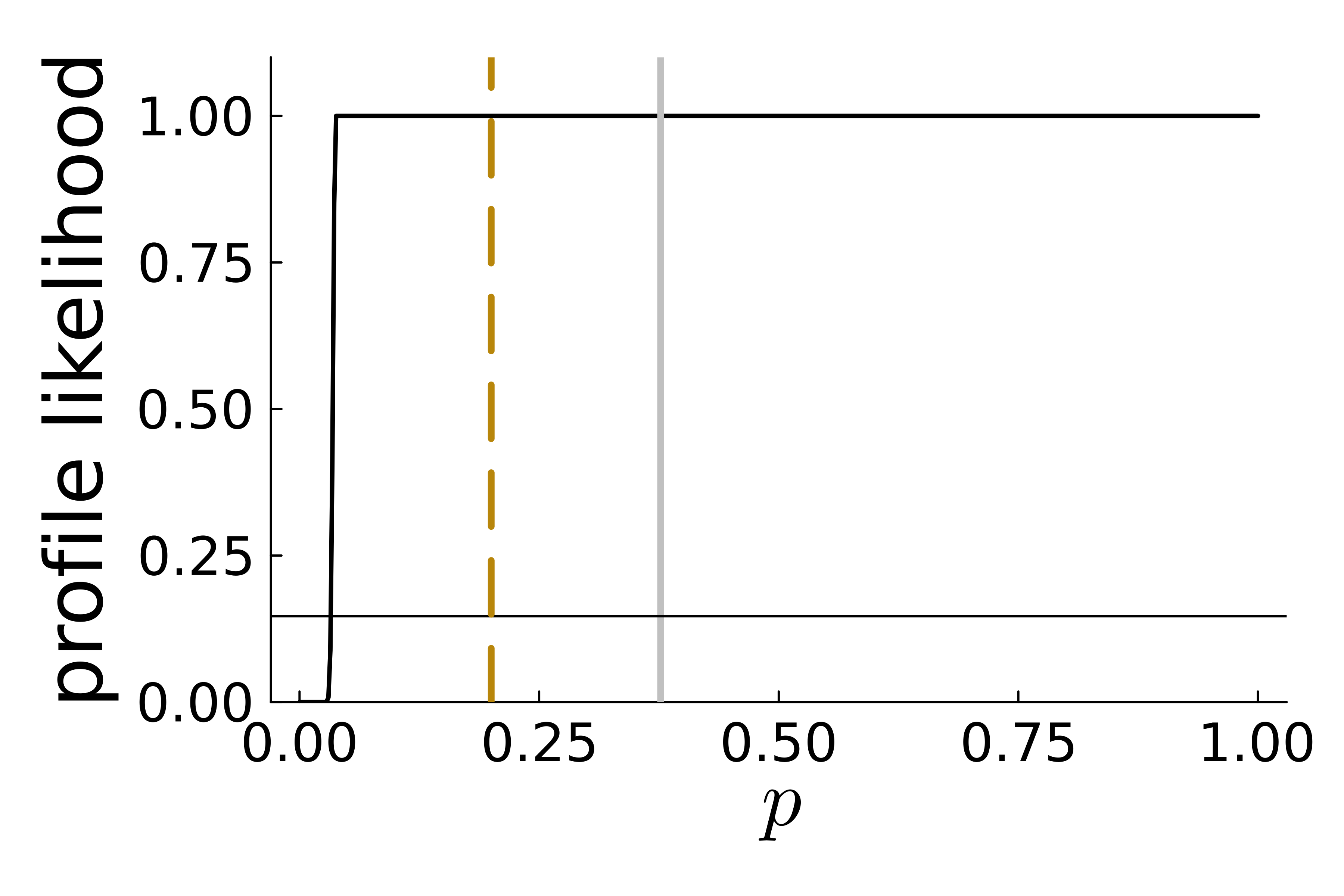}
    \end{subfigure}

    \vspace{0.35cm}
    \begin{subfigure}[t]{0.32\linewidth}
        \includegraphics[width=\linewidth]{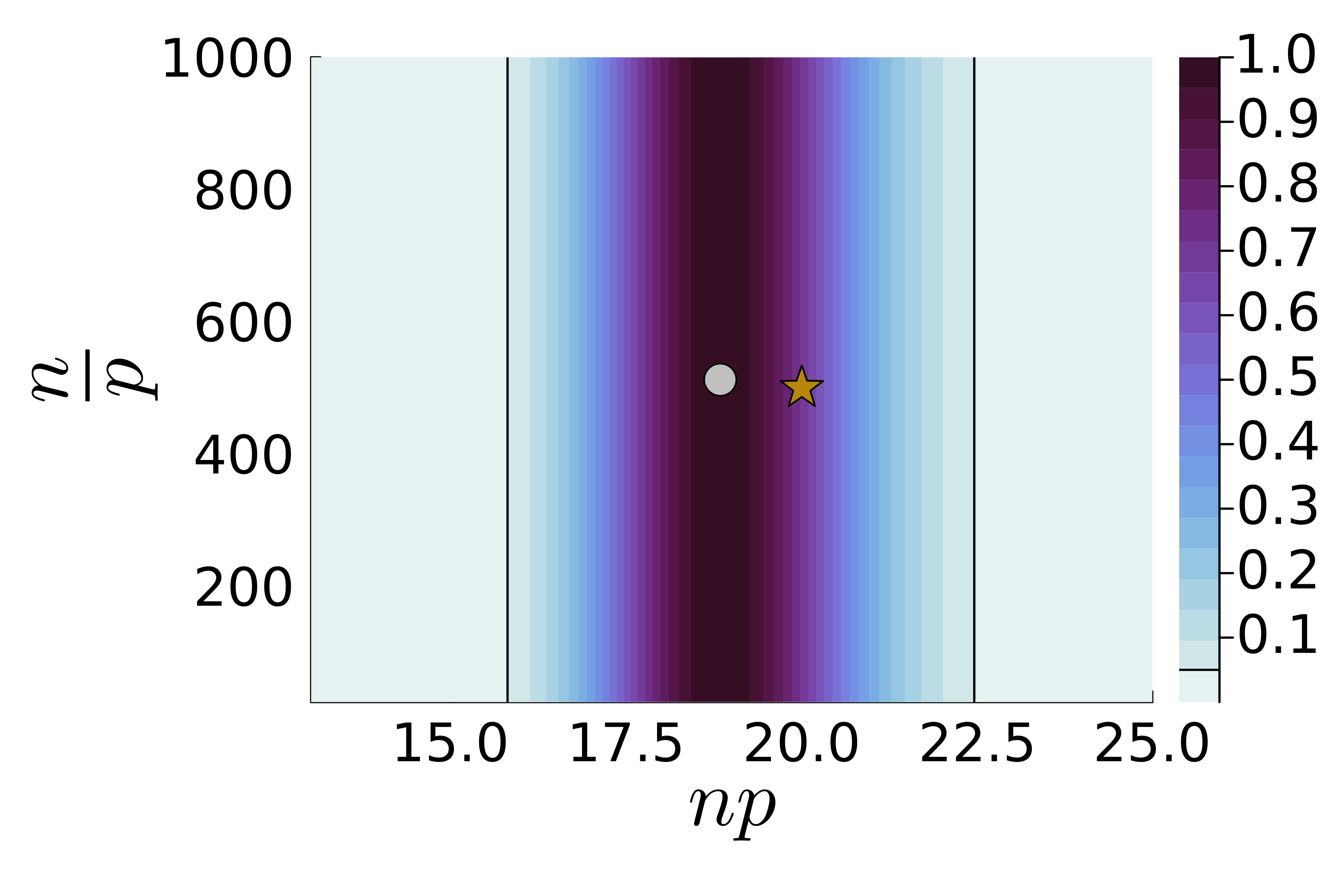}
    \end{subfigure}
    \hfill
    \begin{subfigure}[t]{0.32\linewidth}
        \includegraphics[width=\linewidth]{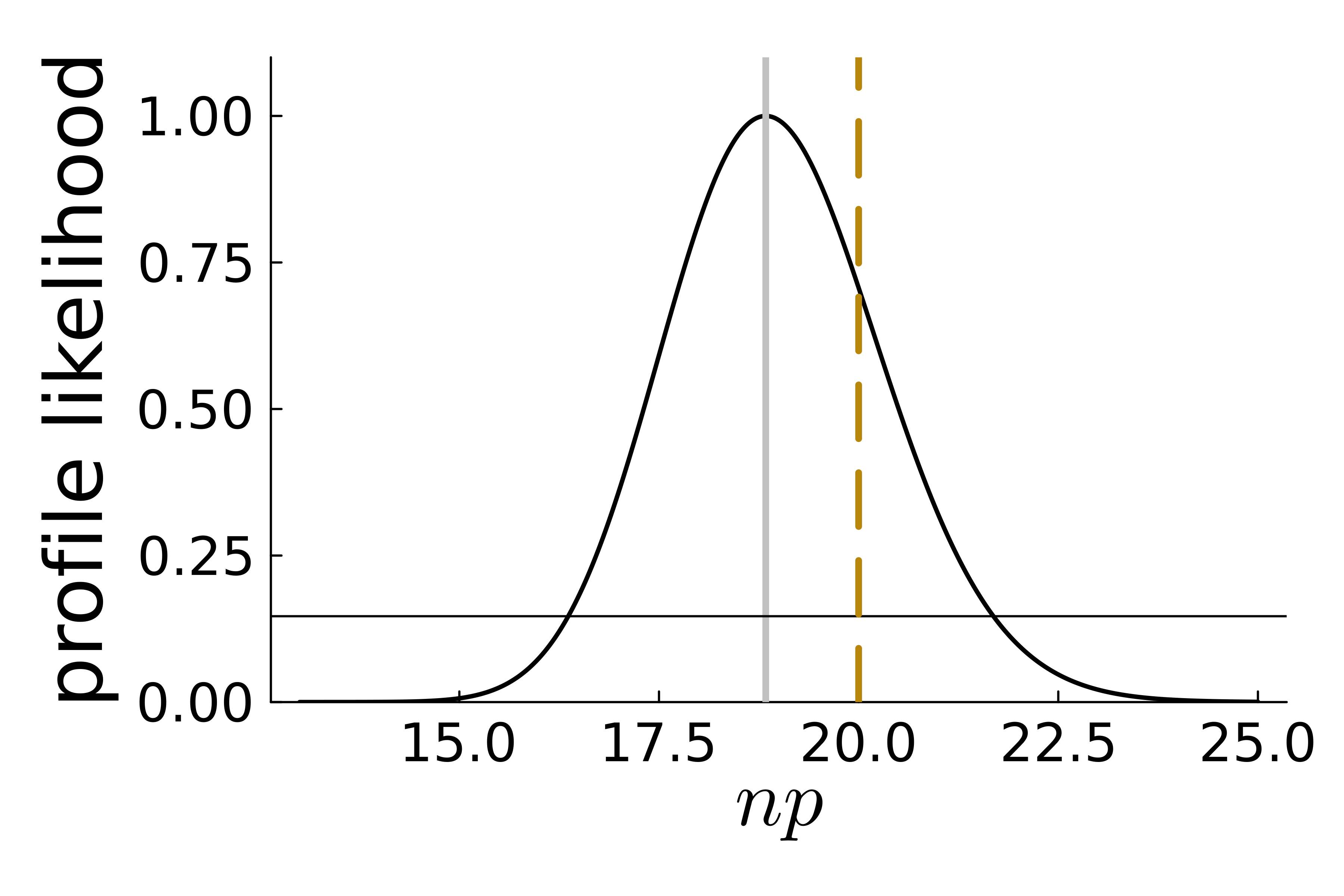}
    \end{subfigure}
    \hfill
    \begin{subfigure}[t]{0.32\linewidth}
        \includegraphics[width=\linewidth]{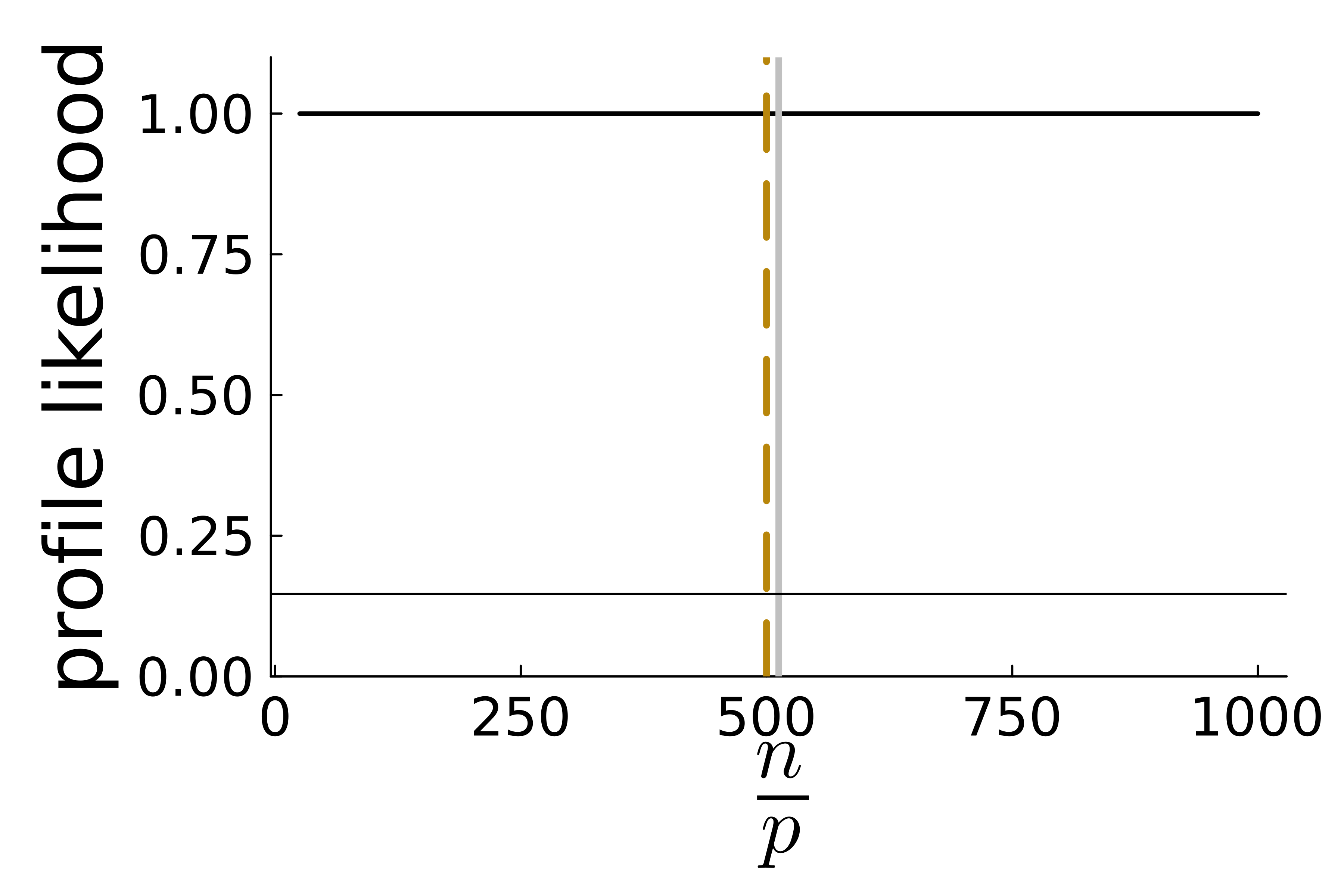}
    \end{subfigure}

    \vspace{0.35cm}
    \begin{subfigure}[t]{0.32\linewidth}
        \includegraphics[width=\linewidth]{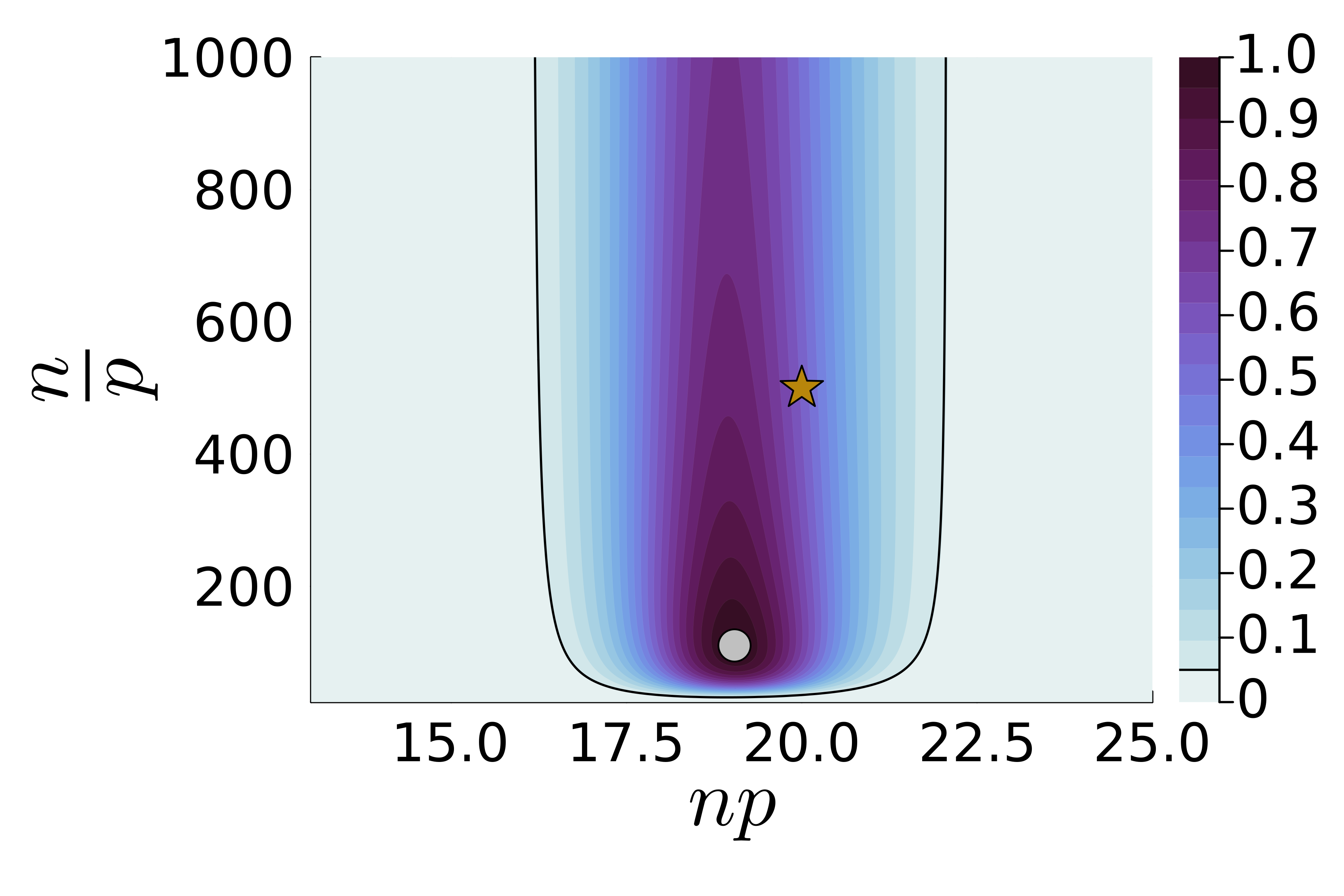}
    \end{subfigure}
    \hfill
    \begin{subfigure}[t]{0.32\linewidth}
        \includegraphics[width=\linewidth]{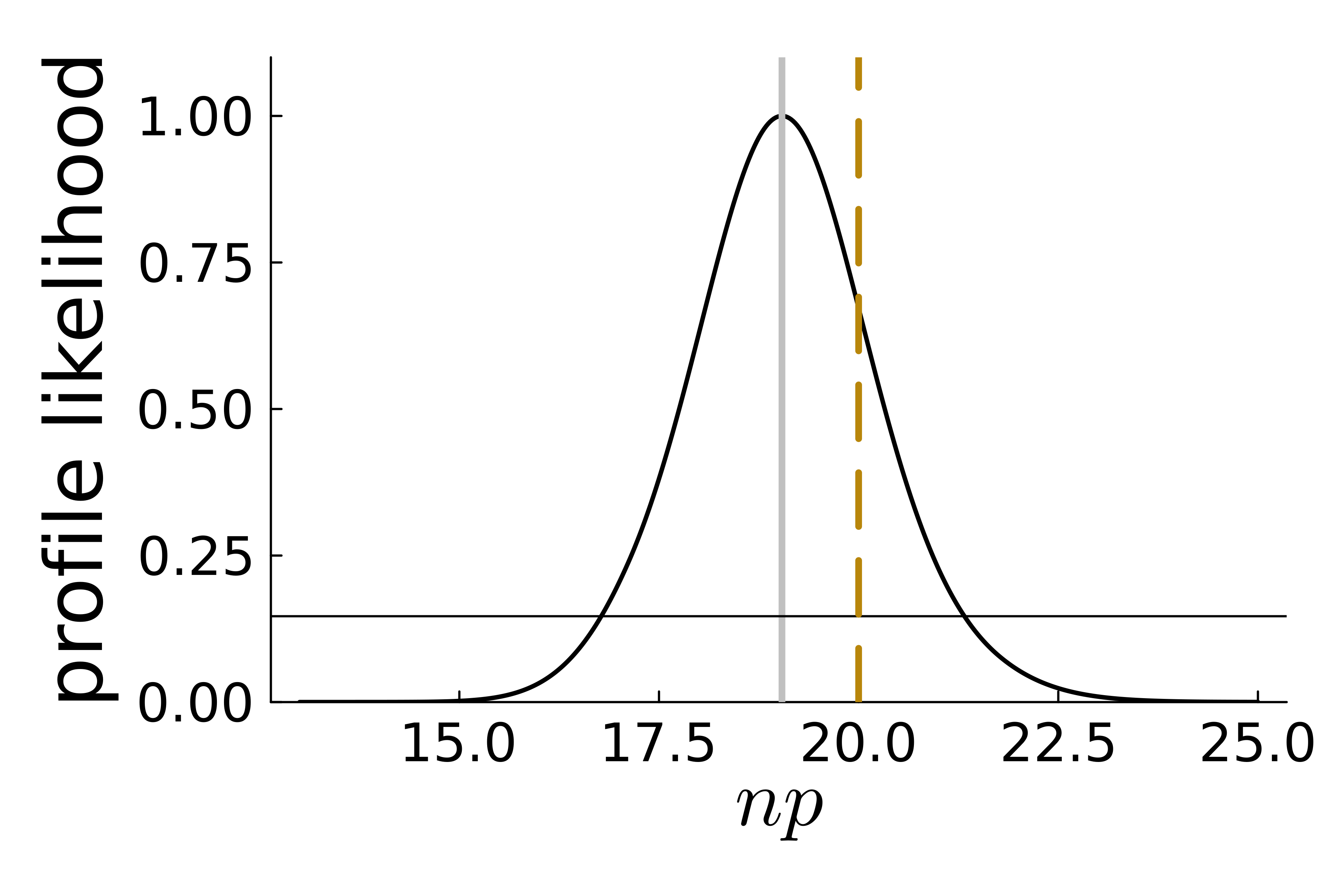}
    \end{subfigure}
    \hfill
    \begin{subfigure}[t]{0.32\linewidth}
        \includegraphics[width=\linewidth]{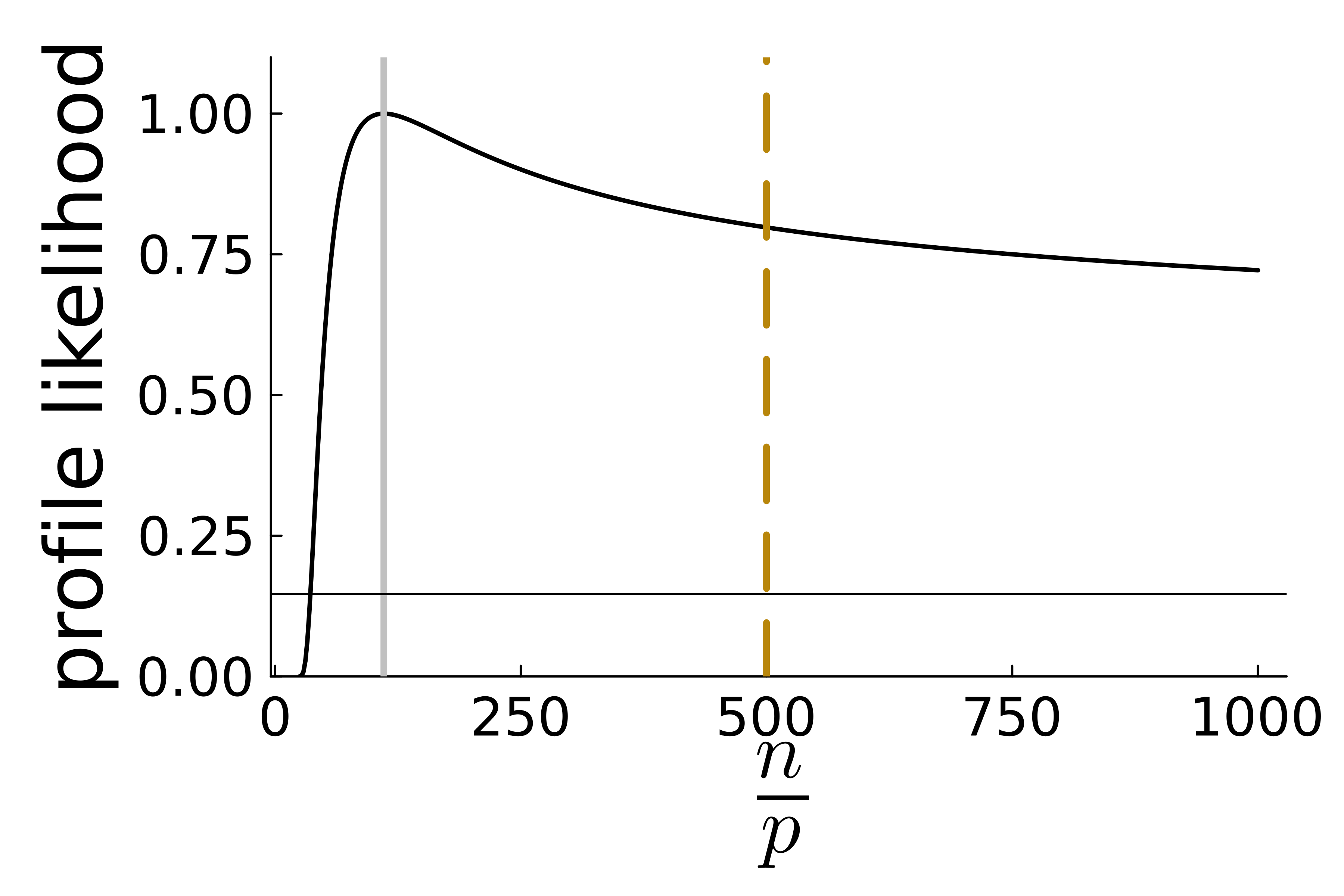}
    \end{subfigure}
    \caption{Numerical illustration of IIR for the normal examples. Top row: Poisson-limit model in the original parameters $(n,p)$, showing the curved likelihood geometry and corresponding one-dimensional profiles. Middle row: the same Poisson-limit model in the transformed coordinates $(np,n/p)$, where $np$ is the identifiable image coordinate and $n/p$ is an invariant null coordinate. Bottom row: non-limit normal model in the same transformed coordinates, where $np$ remains the strong coordinate while $n/p$ is weakly informed rather than exactly non-identifiable.}
    \label{fig:poissonlimit_combined}
\end{figure}

Below we present the results of our numerical analysis of the Poisson-limit and non-limit normal models. We show the likelihood functions in original and reparameterised coordinates in Figure~\ref{fig:poissonlimit_combined}. We used true parameter values of $n = 100$ and $p = 0.2$, a sample size of 10, and bounds of $[0, 500]$ and $[0, 1]$ for $n$ and $p$, respectively. For reproducibility, the data realisation used was [21.9, 22.3, 12.8, 16.4, 16.4, 20.3, 16.2, 20.0, 19.7, 24.4]. This was generated from the non-limit model, i.e. treating the limit model as an approximation at the analysis rather than data generation stage. The analysis can be found in the \href{https://github.com/omaclaren/reparam/}{repository} at \texttt{examples/stat\_model.jl}.

The top row of Figure~\ref{fig:poissonlimit_combined} shows the Poisson-limit likelihood in the original parameterisation. We see a long `banana' shaped likelihood contour in the $(n,p)$ plane, illustrating the existence of (approximately) equivalent parameter combinations lying along a curved relationship. The profile likelihoods for each parameter illustrate the individual non-identifiability of the parameters. The likelihood is completely flat for both parameters, other than where the bounds of the parameter space are reached. Together these results imply that the two parameters are not individually identifiable in this model.

The middle row of Figure~\ref{fig:poissonlimit_combined} shows the Poisson-limit likelihood in reparameterised space. We see the likelihood is constant in the vertical, $\frac{n}{p}$, direction and varies only in the horizontal, $np$, direction. This reflects the identifiability of the parameter combination $np$, and the non-identifiability of the parameter combination $n/p$. Furthermore, this structure means the information concerning the identifiable and non-identifiable parameters is completely separated, in the interior of the domain, as the likelihood factors into a function of $np$ and a constant function of $n/p$ (see e.g. \cite{Aitkin1989,Cox1987}). The profile likelihood for $np$ is a simple Gaussian-like function while the profile likelihood for $n/p$ is completely flat.

The bottom row of Figure~\ref{fig:poissonlimit_combined} shows the likelihood for the non-limit model, defined by \eqref{eq:normal-bin}. This model is structurally identifiable: the mean and variance are now distinct, since $np \neq np(1-p)$ for $p \neq 0$, and we can explicitly invert the auxiliary mapping for $n$ and $p$. Consistent with this, in contrast to the Poisson-limit case, the invariant-subspace calculation gives a trivial invariant null space for this model and hence the identifiable complement is the full space. However, Algorithm~\ref{alg:reparam} still provides useful local practical identifiability information: the SVD of the log-coordinate Jacobian, which is the default starting point of the algorithm, gives one dominant and one weaker right-singular direction. These are not simple monomial directions, but the informed sparse monomial basis selection then gives the (same) interpretable monomial coordinates $np$ and $n/p$, and these are closely aligned in log-space with the dominant and weak right-singular directions. The numerical results hence largely mirror those of the Poisson-limit model and the more strongly identifiable coordinate $np$ and the weaker coordinate $n/p$ are again clearly separated in the reparameterised space. Since, in this case, the separation is not exact, the profile likelihood for $n/p$ is not completely flat, but is weakly curved. Notably, we see clear one-sided practical non-identifiability in $n/p$: the likelihood is much flatter for larger values of $n/p$ than for smaller values, reflecting the fact that the model becomes closer to the Poisson limit for larger values of $n/p$.

\subsection{Repressilator}\label{sec:repressilator-results}
We now consider the repressilator model defined by \eqref{eq:repressilator}. With the Hill coefficient fixed at $n=2.5$, this gives an 18-parameter model with three observed mRNA trajectories. Applying Algorithm~\ref{alg:reparam} at a reference point determined as a maximum likelihood estimate, within the estimation bounds given in the Supplementary Material, gives a Jacobian of rank 15. The local null space is therefore 3-dimensional, and all three null directions pass the invariance test. An additional argument shows that these directions are exactly invariant (not just to first order): multiplying $\beta_i$ and $K_i$ by the same factor $c_i>0$ multiplies the protein trajectory $p_i$ by $c_i$ and leaves the mRNA trajectories unchanged, because the proteins enter the mRNA equations only through $p_i/K_i$ and the protein initial conditions are zero. Thus the invariant image is 15-dimensional.

The default SVD basis separates the directions by local information, but is not itself very interpretable. The sparse monomial basis search, by contrast, recovers a simple mechanistic parameterisation spanning the same invariant image. For the invariant-null side, the sparse basis search selects the three products
$\beta_1K_1, \beta_2K_2, \beta_3K_3$ as an interpretable basis. On the complementary image side, ordering the candidates by simplicity, as on the invariant-null side, recovers the three ratios $K_1/\beta_1, K_2/\beta_2, K_3/\beta_3$, together with the remaining 12 singleton identifiable directions, giving an interpretable sparse basis for the 15-dimensional invariant image. The information-ordered search also selects these three ratios, but combines the remaining parameters into products and ratios. This agrees with Eisenberg and Hayashi, who recovered the ratios $K_i/\beta_i$ by identifying the $(\beta_i,K_i)$ pairs through subset profiling and fitting the resulting profile relationships~\cite{Eisenberg2014}; here the same ratios, together with a sparse complementary null-side basis $\beta_iK_i$, are obtained directly from the invariant-image calculation and sparse-basis search.

As a further check, included in the repository, we repeated the Jacobian calculation and sparse-basis selection at seven other fits along the $K_1/\beta_1$ profile. The rank and the identifiable subspace are unchanged, and the simplicity-ordered search returns the same basis at every fit. The SVD basis varies between fits, however, and so do the products and ratios selected by the information-ordered search, although it always selects the three ratios $K_i/\beta_i$, or their reciprocals.

As indicated above, fixing $n$ is not actually required for the invariant-image calculation. As a supplementary check, included in the repository, we repeat the initial IIR analysis without recomputing the full profile-wise analysis, treating $n$ as an additional positive parameter. The repression terms then contain $(p_i/K_i)^n$, so the auxiliary map is not monomial in the full parameter vector. Nevertheless, the mRNA auxiliary map can be written in terms of log-linear coordinates, with $\log K_i-\log\beta_i$ and $\log n$ appearing as separate transformed parameter combinations. Correspondingly, the IIR calculation gives rank $16$ in $19$ parameters, with the same three-dimensional invariant-null space spanned by $\beta_iK_i$, and with $n$ retained on the image side. Thus IIR can be applied in this particular case where standard differential-algebra methods are not strictly applicable.

To illustrate the likelihood implications, we focus on the representative gene-1 pair $K_1/\beta_1$ and $\beta_1K_1$ in the fixed-$n$ case. We treat this pair as a two-dimensional interest space and compute a two-dimensional profile likelihood over these coordinates while optimising over the remaining 16 nuisance parameters within the wider profile bounds given in the Supplementary Material. Figure~\ref{fig:repressilator_profiles} shows the resulting likelihood in these transformed coordinates (top left), the corresponding accepted region mapped back to the original coordinates (top right), and the one-dimensional profiles for the transformed coordinates (bottom row). The profile likelihood for $K_1/\beta_1$ is bounded and clearly peaked, indicating that this ratio is identifiable in the present setting. Several nuisance parameters reach their bounds along this profile, however, so its shape and the resulting interval depend on those bounds. By contrast, the profile for $\beta_1K_1$ is flat over the chosen range, indicating non-identifiability of this combination. The same qualitative pattern is observed for the corresponding gene-2 and gene-3 pairs (not shown). Thus, while the individual parameters $K_i$ and $\beta_i$ are not separately identifiable, the ratios $K_i/\beta_i$ provide natural identifiable coordinates for the reduced representation.

\begin{figure}[!htbp]
    \centering
    \includegraphics[width=0.85\linewidth]{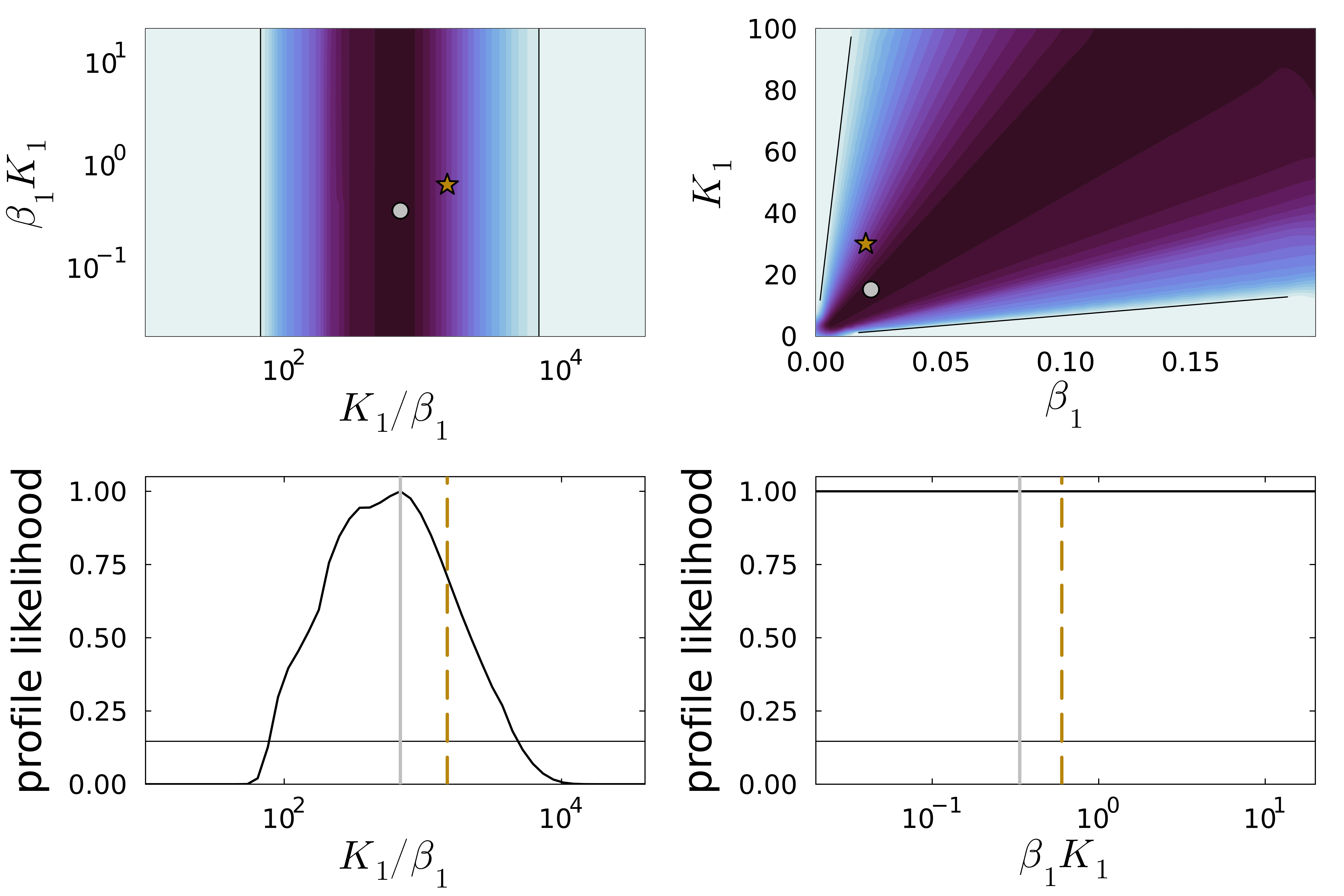}
    \caption{Repressilator likelihood analysis in transformed coordinates, using the representative interest parameter pair $(K_1/\beta_1, \beta_1K_1)$. Top left: joint 2D profile likelihood in IIR coordinates with the remaining 16 parameters profiled out. Top right: the accepted region mapped back to the original $(\beta_1, K_1)$ plane. Bottom row: one-dimensional profiles for the identifiable ratio $K_1/\beta_1$ and the non-identifiable product $\beta_1K_1$. Silver markers indicate gridded maximum-likelihood points, gold markers indicate true values, and black lines indicate the 95\% threshold.}
    \label{fig:repressilator_profiles}
\end{figure}

\begin{figure}[!htbp]
    \centering
    \includegraphics[width=\linewidth]{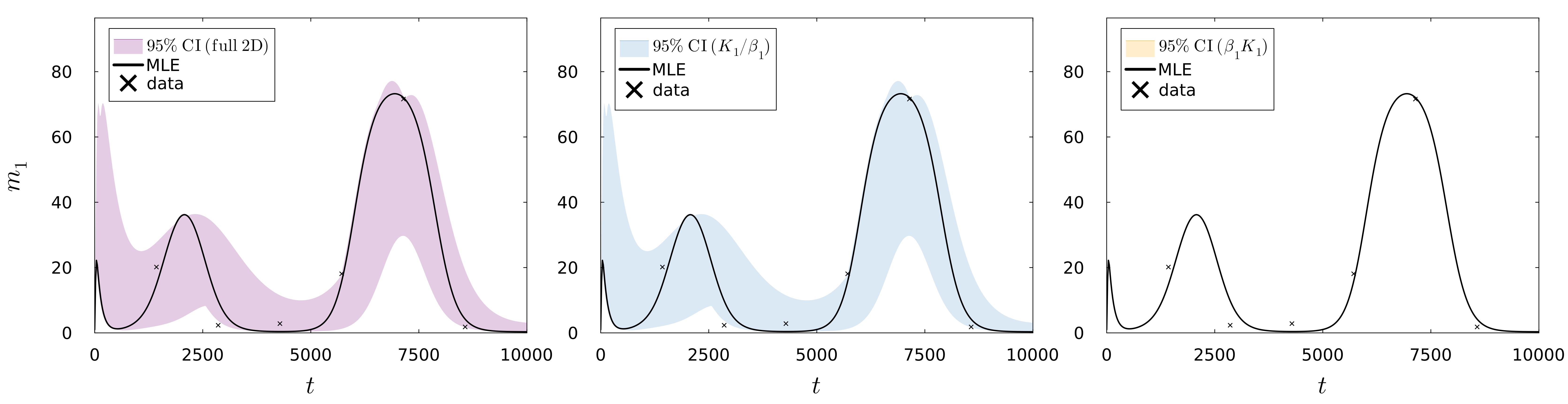}

    \vspace{0.25cm}
    \includegraphics[width=\linewidth]{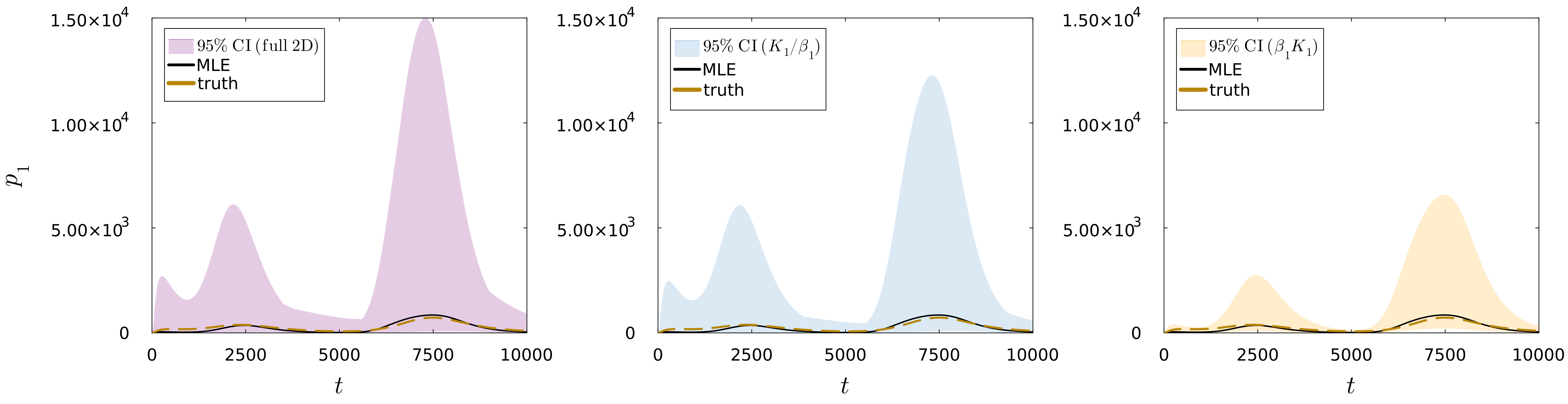}
    \caption{Profile-wise prediction bands for representative observed and unobserved repressilator trajectories, using the interest parameter pair $(K_1/\beta_1,\beta_1K_1)$. Top row: observed mRNA trajectory $m_1$. Bottom row: unobserved protein trajectory $p_1$. Columns show the pushforward of the full accepted two-dimensional profile-likelihood region, the one-dimensional profile over the identifiable ratio $K_1/\beta_1$, and the one-dimensional profile over the non-identifiable product $\beta_1K_1$.}
    \label{fig:repressilator_predictions}
\end{figure}

The top row of Figure~\ref{fig:repressilator_predictions} shows the corresponding profile-wise prediction bands for the observed mRNA trajectory $m_1$. These are obtained by pushing forward (i) the full accepted two-dimensional region in $(K_1/\beta_1,\beta_1K_1)$ space, (ii) the one-dimensional profile over the identifiable coordinate $K_1/\beta_1$, and (iii) the one-dimensional profile over the non-identifiable coordinate $\beta_1K_1$. For visualisation, the contour plot in Figure~\ref{fig:repressilator_profiles} uses the usual two-degree-of-freedom likelihood threshold for the selected two-dimensional interest plane. Since only one of these coordinates is identifiable, this threshold is asymptotically conservative, and a one-degree-of-freedom threshold could be used instead. By contrast, the accepted set used for the prediction pushforward is defined using $df=\rank(D\phi)=15$, corresponding to the effective number of identifiable directions. This follows the PWA interpretation of profile paths as approximating a higher-dimensional accepted region~\cite{Trent2024}. Because these paths are driven by variation in particular parameters or parameter combinations, the resulting prediction bands are not generally uniform-width intervals around the maximum-likelihood prediction; instead, they show where the predictions are more or less sensitive to variation along the chosen profile directions~\cite{Simpson2023}.

The prediction band obtained by varying the identifiable coordinate $K_1/\beta_1$ closely tracks the prediction uncertainty band from the full accepted two-dimensional region. By contrast, profiling only the non-identifiable coordinate $\beta_1K_1$ adds essentially no uncertainty in the observed outputs. Thus, for the observed mRNA trajectories in this example, the band from the identifiable coordinate alone captures most of the uncertainty in the band from the full two-dimensional region, while variation along the non-identifiable direction has essentially no effect on the observations. An important caveat is that the non-identifiable coordinate may have more effect on other, unobserved outputs and on out-of-distribution predictions~\cite{Simpson2024}.

To examine an example of unobserved outputs, the bottom row of Figure~\ref{fig:repressilator_predictions} shows the corresponding profile-wise prediction bands for the protein trajectory $p_1$. Here, in contrast to the mRNA case, variation along the interest parameter $\beta_1K_1$, which is non-identifiable from the mRNA trajectories alone, has a visible effect on the first protein trajectory. This visual effect is consistent with a local augmented-IIR diagnostic: when the auxiliary map is expanded from the three mRNA trajectories to include protein 1, giving $(m_1,m_2,m_3,p_1)$, the Jacobian rank increases from 15 to 16 and the invariant-null dimension drops from 3 to 2. The sparse search shows $\beta_1K_1$ is removed from the null side while $\beta_2K_2$ and $\beta_3K_3$ remain. This indicates that direct measurements of the first protein provide information about $\beta_1K_1$ not available from the mRNA trajectories alone, and illustrates how non-identifiability, observation, and prediction targets interact. Similarly, augmenting the auxiliary map with all three protein trajectories gives a full rank of 18 and removes the invariant-null side entirely, establishing local structural identifiability at the reference point under the augmented observation scenario. The augmented calculation thus finds numerically, without symbolic analysis, that each protein depends on its own product $\beta_iK_i$. The scaling above confirms this: with $K_i/\beta_i$ and the other retained parameters fixed, $p_i$ is proportional to $\sqrt{\beta_iK_i}$. Since the mRNA data do not inform $\beta_1K_1$, the protein bands in Figure~\ref{fig:repressilator_predictions} depend on its profile bounds, and would be unbounded above without them.

\FloatBarrier

\section{Conclusions and Future Work}
We have presented Invariant Image Reparameterisation (IIR), an approach to parameter identifiability analysis that links symbolic reparameterisation ideas with local numerical calculations. The central object in our approach is what we call the invariant image: a basis-independent representation of the parameter combinations associated with the observable behaviour of the model. When the observable behaviour depends on the parameters through a componentwise change of variables followed by a constant lower-dimensional linear reduction, and the reduced map has an injective derivative at a reference point, the symbolic reparameterisation condition has a local numerical form: a Jacobian evaluated at that point gives the row and null-space information sufficient to define the invariant image. Under these conditions, IIR can replace what would otherwise be a symbolic global calculation by an equivalent numerical calculation, up to numerical tolerances. A first-order invariance condition can be used to rule out the minimal case and to extract candidate invariant null directions, including in cases where the resulting image reparameterisation is exact but non-minimal. Keeping all coordinates always gives an exact change of coordinates, while fixing directions that are invariant only to first order gives an approximate reduction. This approach also provides analogous information in settings with practical rather than structural identifiability issues.

Our approach treats the remaining non-uniqueness of reduced parameterisations, whether symbolic or numerical, as a problem of basis choice in the invariant linear null and row spaces, and of coordinate choice in the associated invariant image. The default SVD construction gives an orthonormal reduction matrix with rows ordered by local identifiability, while alternative bases can be chosen for interpretation. In particular, sparse monomial bases can give simpler and more mechanistically meaningful parameter combinations while spanning the same invariant row or invariant null subspaces as the SVD basis. Thus the main output of the method is an invariant image together with possible coordinate systems on that image.

Our examples illustrate IIR for increasing levels of complexity. In the parameterised normal model examples, the IIR analysis of the Poisson-limit case reproduces the expected identifiable combination $np$ and invariant null coordinate $n/p$. In the non-limit case, the model is structurally identifiable, but the same calculations highlight an analogous strong--weak separation in image coordinates, with $np$ strongly informed and $n/p$ weakly informed in a one-sided manner toward the Poisson limit. In the extended Poisson-limit case, IIR analysis shows how a first-stage exact image reparameterisation can give a useful reduction that is non-minimal, and how a subsequent reduction can recover the minimal combination $n_1p_1+n_2p_2$. In the repressilator, the method and sparse basis heuristic recover natural mechanistic ratios and products from a larger 18-parameter nonlinear ODE model. The corresponding profile-wise prediction analysis shows that, for the observed mRNA trajectories, prediction uncertainty is mainly carried by the identifiable ratio $K_1/\beta_1$, while the non-identifiable product $\beta_1K_1$ has little effect on the observed outputs. In contrast, the same non-identifiable coordinate affects an unobserved protein trajectory. An augmented IIR calculation shows that observing the first protein would remove this null direction.

Several directions remain for future work. One is to systematically extend the present largely single-stage framework to more general sequential or multi-stage constructions for cases involving more complex composition of parameter combinations. Another is to develop more systematic methods for choosing interpretable bases within the (linear subspaces associated with the) invariant image, while maintaining general alignment with the most informed directions. A further application of interest would be to biological models where the identifiable combinations are not already known, particularly models with several competing weakly informed combinations. It would also be useful to demonstrate the approach on further model classes within the present scope, such as partial differential equations, as well as to extend it to broader classes of models and auxiliary mappings, including stochastic and simulation-based settings~\cite{Browning2020}, and to investigate more carefully how invariant-image coordinates relate to uncertainty in non-observed quantities and out-of-distribution predictions~\cite{Simpson2024}. For both partial differential equations and stochastic models, the main difficulty is obtaining reliable derivatives of a suitable auxiliary mapping. For partial differential equations, this means differentiating through more complex numerical solvers, and the numerical rank of the Jacobian can depend on the mesh and the observation operator. For stochastic models, the auxiliary mapping should represent the probability distribution of the data, or chosen summaries of it, either exactly or approximately, for example by moment closure or the linear noise approximation~\cite{Schnoerr2017}, and such approximations can introduce further error.

\section{Acknowledgements}
\noindent
Parts of this work were carried out at the MATRIX Workshop on Parameter Identifiability in Mathematical Biology (2024). We thank MATRIX for funding this workshop and the participants for stimulating discussions. We thank the Associate Editor and anonymous referees of an earlier version of this work for constructive comments that substantially improved the manuscript. The authors wish to acknowledge use of the eResearch Infrastructure Platform hosted by the Crown company, Research and Education Advanced Network New Zealand (REANNZ) Ltd., and funded by the Ministry of Business, Innovation \& Employment. URL: \url{https://www.reannz.co.nz}. MJS was supported by the Australian Research Council Centre of Excellence for the Mathematical Analysis of Cellular Systems grant no. CE230100001 (MJS). Large Language Models (LLMs), including Claude, Gemini, and ChatGPT, together with coding-assistance tools such as GitHub Copilot, Codex, Claude Code, and Pi, were used during manuscript refinement and code development for drafting and editorial suggestions, assessing technical arguments, code refactoring, and implementation assistance. All suggestions from these tools were critically evaluated before incorporation. All technical content, proofs, algorithms, analyses, and final wording were checked and approved by the authors, who take responsibility for the manuscript and accompanying code.

\section{Author Contributions}
All authors contributed to the review and editing of the manuscript. OJM wrote the original draft with contributions from MJS. Both OJM and MJS contributed to the conceptualisation of the article, choice of examples and design of numerical experiments. OJM developed the method details and proofs, with assistance from RN, and wrote the final code, drawing on PWA-related code and code review by JAT, and on initial example implementation work by JR.

\section{Competing interests}
The authors declare that they have no competing interests.

\section{Data and materials availability}
All data and code needed to evaluate the conclusions in the paper are present in the paper, Supplementary Material, and the GitHub repository: \href{https://github.com/omaclaren/reparam}{\texttt{https://github.com/omaclaren/reparam}}.

\bibliographystyle{siamplain}
\bibliography{references}

\clearpage
\section*{Supplementary Materials: Invariant Image Reparameterisation}
\setcounter{section}{0}
\setcounter{subsection}{0}
\setcounter{equation}{0}
\setcounter{figure}{0}
\setcounter{table}{0}
\setcounter{algorithm}{0}
\renewcommand{\thesection}{SM\arabic{section}}
\renewcommand{\thefigure}{SM\arabic{figure}}
\renewcommand{\thetable}{SM\arabic{table}}
\def\theHsection{SM.\arabic{section}}
\def\theHfigure{SM.\arabic{figure}}
\def\theHtable{SM.\arabic{table}}
\def\theHalgorithm{\thealgorithm}

\section{Additional method details}
Here we present additional arguments and results supporting the main text.

\subsection{Fisher information analysis}
As mentioned in the main text, the auxiliary mapping and observed Fisher information can provide the same rank information when defined on the same grid level, under the conditions given below. Here we assume this is the solution grid level. We can also work with the finite data observed Fisher information, but this will not generally exactly recover the structural identifiability results due to additional discretisation. We can also work with the expected Fisher information, but prefer to consider the observed Fisher information as it is more directly related to the bridge between practical identifiability and structural identifiability. The link between identification and expected Fisher information for exponential family models is discussed in~\cite{Cole2020}; an early reference is~\cite{Catchpole1997}.

To obtain the required result for the observed Fisher information, we first consider the transformation of the observed Fisher information under a change of parameters. Given the observed Fisher information in terms of the data distribution parameters, $\phi$, and the Jacobian of the auxiliary mapping, $\phi_{\theta}(\theta)$, using subscripts to denote derivatives for compactness and taking $l(\phi;y)$ to denote the negative log-likelihood, we can write the observed Fisher information in terms of the mechanistic model parameters, $\theta$, as
\begin{equation}
\begin{aligned}
    \mathcal{J}(\theta) &= \phi_{\theta}(\theta)^T l_{\phi\phi}(\phi(\theta);y) \phi_{\theta}(\theta) + (l_{\phi}(\phi(\theta);y) \otimes I_{p} )\phi_{\theta\theta}(\theta)\\
    &=\phi_{\theta}(\theta)^T \mathcal{J}(\phi(\theta))\phi_{\theta}(\theta) + (l_{\phi}(\phi(\theta);y) \otimes I_{p} )\phi_{\theta\theta}(\theta),
\end{aligned}
\end{equation}
where $\otimes$ is the Kronecker matrix product, so that the second term is $\sum_{a=1}^q l_{\phi_a}\phi_{a,\theta\theta}$, and where the above follows from the chain rule for matrix derivatives~\cite{Magnus2019}; see also~\cite{Barndorff2014, Pace1997}.

The data-distribution parameters $\phi$ need not vary independently within the model. For the argument below, we assume a local minimal image reparameterisation $\phi=\tilde{\phi}\circ\psi$, with $\psi_\theta$ of full row rank and $\tilde{\phi}_\psi$ of full column rank. At an interior maximum likelihood estimate $\hat{\theta}$, the negative log-likelihood has zero gradient with respect to $\theta$. Full row rank of $\psi_\theta$ then implies that the gradient with respect to $\psi$ is also zero. This enables us to drop the term involving this gradient when applying the chain rule in these coordinates, giving
\begin{equation}
\begin{aligned}
\mathcal{J}(\hat{\theta}) = \psi_{\theta}(\hat{\theta})^T \mathcal{J}(\hat{\psi})\psi_{\theta}(\hat{\theta}),
\end{aligned}
\end{equation}
where $\hat{\psi}=\psi(\hat{\theta})$ and $\mathcal{J}(\hat{\psi})$ is the observed information for $l(\tilde{\phi}(\psi);y)$. The transformation law has the same form as for one-to-one reparameterisations~\cite{Pace1997, Barndorff2014}. The local reparameterisation used in this argument need not be the one returned by IIR.

The observed information in these reduced coordinates includes both terms from the chain rule,
\begin{equation}
\mathcal{J}(\hat{\psi}) = \tilde{\phi}_\psi^T l_{\phi\phi}\tilde{\phi}_\psi
+\sum_{a=1}^q l_{\phi_a}\tilde{\phi}_{a,\psi\psi},
\end{equation}
with all terms evaluated at the fit for the fixed data $y$.

Next, we assume that $\mathcal{J}(\hat{\psi})$ is positive definite, i.e., $\hat{\psi}$ is a non-degenerate local maximum of the likelihood in the reduced coordinates. Hence we have $\mathcal{J}(\hat{\psi})=L^TL$ for some invertible square matrix $L$, and so
\begin{equation*}
\begin{aligned}
    \mathcal{J}(\hat{\theta}) &= \psi_{\theta}(\hat{\theta})^T \mathcal{J}(\hat{\psi})\psi_{\theta}(\hat{\theta})\\
    &= \psi_{\theta}(\hat{\theta})^T L^T L \psi_{\theta}(\hat{\theta})\\
    &= (L \psi_{\theta}(\hat{\theta}))^T(L \psi_{\theta}(\hat{\theta})).
\end{aligned}
\end{equation*}
Here $\phi_\theta=\tilde{\phi}_\psi\psi_\theta$. Since $\text{rank}(A^TA) = \text{rank}(A)$ for any matrix $A$, and $\text{rank}(BC) = \text{rank}(C)$ when $B$ has full column rank, it follows directly that
\begin{equation*}
\begin{aligned}
    \text{rank}(\mathcal{J}(\hat{\theta})) &= \text{rank}(\psi_\theta(\hat{\theta})) = \text{rank}(\phi_\theta(\hat{\theta})),\\
    \nullspace\mathcal{J}(\hat{\theta}) &= \nullspace\psi_\theta(\hat{\theta}) = \nullspace\phi_\theta(\hat{\theta}).
\end{aligned}
\end{equation*}
This correspondence concerns the rank and null space, rather than the eigenvalues or a particular choice of basis. For comparison, the expected information satisfies $I_\theta=\phi_\theta^T I_\phi\phi_\theta$ under the usual regularity conditions, with the same rank and null space as $\phi_\theta$ when $I_\phi$ is positive definite~\cite{Catchpole1997}.

In the case of spatial or temporal data, the data distribution parameters (exhaustive summaries) will generally be given by the solution at the fine grid level. If the Fisher information is only defined at the observation grid level then we would need to use $\phi_{\text{obs}} = B_{\text{obs}}\phi$ instead of $\phi$ in the above. If $B_{\text{obs}}$ removes directions in the image of $\phi_\theta$, the rank can decrease and certain parameter effects become unobservable at the discretisation level. This implies that some parameters that are structurally identifiable at the fine grid level may appear practically non-identifiable at the observation grid level. Hence equivalence between the auxiliary mapping and observed Fisher information formulations requires a consistent choice of grid level.

\subsection{First-order extraction of invariant null directions}
The exact single-reference-point IIR calculation follows from the form of the parameter transformation in the main text, when the reduced map has an injective derivative at the reference point. In general, however, the full local null space of $D_{\theta^*}\phi_*(\hat{\theta}^*)$ need not be invariant: in non-minimal image cases only a proper subspace of it may remain constant. Here we give the first-order calculation used to extract candidate invariant null directions.

Let $\theta^*=f(\theta)$ and let $\phi_*(\theta^*)=\phi(f^{-1}(\theta^*))$. Suppose $\hat\alpha$ is a local null vector at a reference point $\hat\theta^*$, so
\begin{equation}
D_{\theta^*}\phi_*(\hat\theta^*)\hat\alpha=0.
\end{equation}
A necessary condition for $\hat\alpha$ to define a constant null direction is that it should remain a null direction under small perturbations of the transformed parameters. Expanding the Jacobian gives
\begin{equation}
D_{\theta^*}\phi_*(\hat\theta^*+\delta)
=
D_{\theta^*}\phi_*(\hat\theta^*)+
\sum_i H_i(\hat\theta^*)\delta_i+O(\|\delta\|^2),
\end{equation}
where
\begin{equation}
H_i(\hat\theta^*)=D_{\theta^*_i}D_{\theta^*}\phi_*(\hat\theta^*)\in\mathbb{R}^{q\times p}
\end{equation}
is the derivative of the Jacobian with respect to the $i$th transformed parameter. Dropping the higher-order term and requiring the perturbed null condition for all small $\delta$ gives the necessary first-order condition
\begin{equation}
H_i(\hat\theta^*)\hat\alpha=0,
\qquad i=1,\dots,p.
\label{eq:supp-first-order-condition}
\end{equation}
Failure of \eqref{eq:supp-first-order-condition} rules out a constant null direction at first order. Passing the condition at a single point does not, by itself, show that a direction is exactly constant, even when the representation $\phi_*(\theta^*)=\bar\phi(A\theta^*)$ with $A$ constant holds. For example, for $\phi_*(\theta^*)=\theta^*_1+(\theta^*_2)^3$ at $\theta^*_2=0$, the direction $(0,1)^T$ passes the condition but is not a null direction elsewhere. Directions that pass are therefore candidates, to be checked further, for example analytically as in the extended example below.

To compute the candidate subspace, first take the SVD
\begin{equation}
D_{\theta^*}\phi_*(\hat\theta^*)=U\Sigma V^T,
\end{equation}
and write $V=\begin{bmatrix}V_r & V_0\end{bmatrix}$ where $V_0$ spans the local null space. Any local null vector can be written as $\hat\alpha=V_0c$. Combining \eqref{eq:supp-first-order-condition} for all parameters gives
\begin{equation}
\begin{bmatrix}
H_1(\hat\theta^*)V_0\\
H_2(\hat\theta^*)V_0\\
\vdots\\
H_p(\hat\theta^*)V_0
\end{bmatrix}C_0
=MC_0=0.
\label{eq:supp-invariance-stack}
\end{equation}
Here the columns of $C_0$ are coefficient vectors for those local null
directions that satisfy the first-order invariance condition. Computationally, the stacked products $H_i(\hat\theta^*)V_0$ need not be formed by constructing the full Hessian tensor. In the implementation we differentiate the Jacobian--null-basis product
$\theta^* \mapsto D_{\theta^*}\phi_*(\theta^*)V_0$ and reshape the result to obtain the blocks $H_i(\hat\theta^*)V_0$.

A default choice for $C_0$ is given by the right singular vectors associated with zero singular values of the stacked matrix; denoting these vectors by $V_{M0}$, a basis for the candidate invariant null directions is given by
\begin{equation}
N=V_0V_{M0}.
\end{equation}
If $N$ has the same rank as $V_0$, every local null direction passes the test. If $N$ has smaller rank, the local null directions outside $N$ are not constant, so no minimal reparameterisation of this form exists; when the directions in $N$ are invariant, the result is an exact but non-minimal image reparameterisation. A canonical orthonormal complement is then obtained by concatenating these column bases,
\begin{equation}
N_\perp=\begin{bmatrix} V_r & V_0V_{Mr}\end{bmatrix},
\end{equation}
where the columns of $V_{Mr}$ are the right singular vectors of $M$ associated with nonzero singular values. The default SVD reduction matrix is $A_{\mathrm{SVD}}=N_\perp^T$.

\subsection{Sparse monomial basis selection}
\label{sec:sparse-basis}

The SVD reduction matrix is orthonormal, but its rows are typically dense linear combinations of log-parameters, corresponding to monomial combinations with non-integer, non-sparse exponents. These are generally hard to interpret mechanistically. We therefore search for vectors of integer exponents that are sparse, lie close to either the invariant-null $\Span (N)$ or complementary image side $\Span (N_{\perp})$ and, for that side, carry as much local information as possible.

The algorithm we use here is a greedy selection over a finite dictionary of candidate integer exponent vectors. The two sides of the invariant image use the same dictionary and the same subspace and rank filters, but differ in how candidates are scored.

\subsubsection{Subspace filter}
Let $Q$ be an orthonormal basis for the target subspace. The projection
residual
\begin{equation}
  r(v) = \frac{\lVert (I - QQ^{\top})\,v \rVert_{2}}{\lVert v \rVert_{2}}
\end{equation}
measures how far a candidate $v$ lies from the target subspace. We
discard candidates whose residual exceeds a small numerical tolerance. Duplicates up to overall sign and common integer factors are also removed, i.e. parallel vectors.

\subsubsection{Rank filter}
A retained candidate is added to the selected set only if it strictly increases the numerical rank of the selected set's projection into the target subspace. This prevents the algorithm from selecting near-duplicate directions and provides a stopping criterion: the selected basis is complete when its rank matches the target dimension.

\subsubsection{Image-side score}
On the image side, we score candidates based on how much new (local) information about the auxiliary-map output a candidate carries, given the candidates already selected. Geometrically, the auxiliary-map Jacobian $J_{\ast} = D_{\theta^{\ast}} \phi_{\ast}(\hat{\theta}^{\ast})$ maps a normalised direction $\hat{v}$ in log-parameter space to a sensitivity vector $J_{\ast}\hat{v}$ in auxiliary-map output space. A direction with a long sensitivity vector produces a large local change in the output. A direction with sensitivity vector lying in the span of already-selected sensitivity vectors produces only changes already captured by the existing basis, and so carries no new information.

First, the unconditional score assigned is the squared length of the sensitivity vector,
\begin{equation}
  s(v \mid \emptyset)
  = \lVert J_{\ast} \hat{v} \rVert_{2}^{2}.
\end{equation}
Then, if $B = [\hat{b}_{1}, \dots, \hat{b}_{k}]$ is the matrix of the already-selected normalised directions and $J_{\ast} B$ the matrix of the associated sensitivity
vectors, the conditional score assigned is the squared length of $J_{\ast}\hat{v}$
after projecting out the part already explained by $J_{\ast} B$:
\begin{equation}
  s(v \mid B)
  = \bigl\lVert J_{\ast} \hat{v}
    - P_{J_{\ast} B}\, J_{\ast} \hat{v}
  \bigr\rVert_{2}^{2},
\end{equation}
where $P_{J_{\ast} B}$ is the orthogonal projector onto
$\Span(J_{\ast} B)$.

We normalise scores by the largest singular value of $J_{\ast}$ squared, which does not affect the ordering. Retained candidates with projection residual below a residual tolerance are searched greedily: at each step, the rank-increasing candidate with the highest conditional score is selected.

\subsubsection{Null-side score}
For candidates in the invariant-null subspace, $J_{\ast}\hat{v} = 0$ by construction and so the sensitivity vector vanishes. Hence the score above cannot discriminate between candidates. We instead drive selection by `simplicity'. This is defined by ordering candidates first by support size (the number of nonzero exponents), then by $\ell_1$-norm, then by $\ell_\infty$-norm, then by projection residual $r(v)$ (if further tie-break is needed).

We accept rank-increasing candidates greedily in this order. Algorithm~\ref{alg:sparse-basis} summarises the procedure.

\begin{algorithm}[H]
\caption{Sparse monomial basis selection}
\label{alg:sparse-basis}
\begin{algorithmic}[1]
\Input Target basis $Q$ for $\Span(N_{\perp})$ or $\Span(N)$; dictionary $D$ of integer exponent vectors; residual tolerance $\tau$; numerical rank tolerance; Jacobian $J_{\ast}$ for the $\Span(N_{\perp})$ case.
\Output Ordered sparse basis $B$.
\State Deduplicate $D$ up to overall sign and common integer factors.
\State Compute residual $r(v)$ for each $v \in D$, and retain candidates with $r(v) \le \tau$.
\State $B \gets \emptyset$;\quad current rank $\gets 0$.
\If{target is $\Span(N_{\perp})$}
    \While{current rank $< \rank(Q)$}
      \State Let $S$ be the retained candidates that strictly increase numerical rank of $B$.
      \If{$S=\emptyset$} \State \textbf{break} \EndIf
      \State Select from $S$ the candidate with highest $s(v \mid B)$.
      \State Append the selected candidate to $B$; update current rank.
    \EndWhile
\Else { // target is $\Span(N)$}
    \For{retained candidates in order $(\mathrm{support}, \ell_1, \ell_\infty, r(v))$}
      \If{the candidate strictly increases numerical rank of $B$}
        \State Append the selected candidate to $B$; update current rank.
        \If{current rank $= \rank(Q)$} \State \textbf{break} \EndIf
      \EndIf
    \EndFor
\EndIf
\State \Return $B$.
\end{algorithmic}
\end{algorithm}

The example-specific dictionary bounds, residual tolerances, and rank thresholds used for the repressilator calculation are given below.

\subsection{Augmented observation maps}\label{sec:supp-augmented-maps}
Let the columns of $V_0$ span the local null space of the Jacobian $J$ of the current observation map, in the transformed coordinates $\theta^*$, and let $q(\theta^*)$ be a vector of predictions or candidate measurements with Jacobian $Dq$. The columns of $Dq\,V_0$ are the first-order changes in $q$ along the null directions, so a null direction $V_0c$ leaves $q$ unchanged to first order exactly when $Dq\,V_0c=0$. These directions form the subspace $V_0\nullspace(Dq\,V_0)=\{V_0c : Dq\,V_0c=0\}$. This subspace is the null space of the augmented Jacobian obtained by stacking $J$ and $Dq$, since a vector is in that null space exactly when it is a null direction $V_0c$ of $J$ with $Dq\,V_0c=0$. Since $V_0$ has full column rank, the subspace has dimension $k-\rank(Dq\,V_0)$, where $k$ is the number of columns of $V_0$. Adding $q$ to the observation map therefore removes $\rank(Dq\,V_0)$ null directions and, by the rank--nullity theorem, increases the rank of the Jacobian by the same amount. In particular, $Dq\,V_0=0$ exactly when the rows of $Dq$ lie in the row space of $J$, so that, to first order, $q$ depends only on the identifiable combinations.

\section{Parameterised normal model diagnostics}
The main text contains the analytical Poisson-limit calculation and summarises the extended Poisson-limit calculation. Here we give the additional algebra for the extended case, where the first IIR calculation gives an exact but non-minimal image reparameterisation.

\subsection{Extended Poisson-limit details}
For the original parameterisation of the extended model, the auxiliary mapping is given by
\begin{equation}
\begin{aligned}
    \phi(n_1, p_1, n_2, p_2) =
\begin{bmatrix}
    n_1p_1 + n_2p_2\\
    n_1p_1 + n_2p_2
\end{bmatrix}.
\end{aligned}
\end{equation}
This model is clearly structurally non-identifiable, as only the sum of products $n_1p_1 + n_2p_2$ determines the output of the auxiliary mapping.

The Jacobian of this mapping is given by
\begin{equation}
\begin{aligned}
    D\phi(n_1, p_1, n_2, p_2) = \begin{bmatrix}
    p_1 & n_1 & p_2 & n_2\\
    p_1 & n_1 & p_2 & n_2
\end{bmatrix}.
\end{aligned}
\end{equation}
This has rank one and a null space of dimension three. One convenient symbolic basis (non-orthogonal) for the null space is given by the vectors
\begin{equation}
\begin{aligned}
    \alpha_1(n_1, p_1, n_2, p_2) = \begin{bmatrix}
    n_1\\
    -p_1\\
    0\\
    0
\end{bmatrix},
\qquad
\alpha_2(n_1, p_1, n_2, p_2) = \begin{bmatrix}
    0\\
    0\\
    n_2\\
    -p_2
\end{bmatrix},
\qquad
\alpha_3(n_1, p_1, n_2, p_2) = \begin{bmatrix}
    p_2\\
    0\\
    -p_1\\
    0
\end{bmatrix}.
\end{aligned}
\end{equation}
Solving the system of partial differential equations
\begin{equation}
\begin{aligned}
    D\psi \alpha_i = 0, \qquad i=1,2,3,
\end{aligned}
\end{equation}
gives, as one solution, an identifiable parameterisation $\psi(n_1, p_1, n_2, p_2) = n_1p_1 + n_2p_2$, as expected.

For IIR, consider the extended Poisson limit model in log-transformed coordinates, defined by $n_i^* = \log n_i$ and $p_i^* = \log p_i$ for $i=1,2$. The Jacobian of the auxiliary mapping in these coordinates is given by
\begin{equation}
\begin{aligned}
    D\phi_*(n_1^*, p_1^*, n_2^*, p_2^*) = \begin{bmatrix}
    \exp(n_1^* + p_1^*) & \exp(n_1^* + p_1^*) & \exp(n_2^* + p_2^*) & \exp(n_2^* + p_2^*)\\
    \exp(n_1^* + p_1^*) & \exp(n_1^* + p_1^*) & \exp(n_2^* + p_2^*) & \exp(n_2^* + p_2^*)
    \end{bmatrix}.
\end{aligned}
\end{equation}
The null space of this admits a basis composed of two constant vectors:
\begin{equation}
\begin{aligned}
    \beta_1 = \begin{bmatrix}
    1\\
    -1\\
    0\\
    0
\end{bmatrix},
\qquad
\beta_2 = \begin{bmatrix}
    0\\
    0\\
    1\\
    -1
\end{bmatrix},
\end{aligned}
\end{equation}
and a third vector that depends on the parameters:
\begin{equation}
\begin{aligned}
\beta_3(n_1^*, p_1^*, n_2^*, p_2^*) = \begin{bmatrix}
    \exp(n_2^* + p_2^*)\\
    0\\
    -\exp(n_1^* + p_1^*)\\
    0
\end{bmatrix}.
\end{aligned}
\end{equation}
Thus the full null space of the Jacobian is not invariant, but the two-dimensional subspace spanned by the constant vectors is invariant. The orthogonal complement of this invariant subspace is two-dimensional and spanned by the constant row vectors
\begin{equation}
\begin{aligned}
    \begin{bmatrix}1 & 1 & 0 & 0\end{bmatrix},
    \qquad
    \begin{bmatrix}0 & 0 & 1 & 1\end{bmatrix}.
\end{aligned}
\end{equation}
Thus we can recover a non-minimal image reparameterisation with $A$ constructed from the two constant row vectors spanning the orthogonal complement of the invariant subspace, giving
\begin{equation}
\begin{aligned}
\eta = A\begin{bmatrix}n_1^*\\
p_1^*\\
n_2^*\\
p_2^*\end{bmatrix},
\qquad
A=
\begin{bmatrix}
1 & 1 & 0 & 0\\
0 & 0 & 1 & 1
\end{bmatrix}.
\end{aligned}
\end{equation}
In reduced transformed coordinates, the auxiliary mapping is
\begin{equation}
\bar{\phi}(\eta)
=
\begin{bmatrix}
\exp(\eta_1)+\exp(\eta_2)\\
\exp(\eta_1)+\exp(\eta_2)
\end{bmatrix}.
\end{equation}
Applying $g = \exp$ to transform the reduced coordinates to monomial form gives the coordinates
\begin{equation}
\begin{aligned}
    \psi(n_1, p_1, n_2, p_2) &= \exp\left(\begin{bmatrix}
    1 & 1 & 0 & 0\\
    0 & 0 & 1 & 1
    \end{bmatrix}\begin{bmatrix}
    n_1^*\\
    p_1^*\\
    n_2^*\\
    p_2^*
    \end{bmatrix}\right)\\
&= \begin{bmatrix}
    \exp(\eta_1)\\
    \exp(\eta_2)
\end{bmatrix}
=
\begin{bmatrix}
n_1p_1\\
n_2p_2
\end{bmatrix},
\end{aligned}
\end{equation}
and so the auxiliary mapping in these coordinates is given by
\begin{equation}
\begin{aligned}
    \tilde{\phi}(\psi) =
\begin{bmatrix}
    \psi_1 + \psi_2\\
    \psi_1 + \psi_2
\end{bmatrix}.
\end{aligned}
\end{equation}
Here we can see from the analytical calculation that one of the null space vectors is parameter-dependent, but this is not available in a single-point numerical calculation. This is where the first-order invariance calculation is useful, as it both indicates the existence of a lower-dimensional invariant subspace and determines this subspace. From the Hessian slices
\begin{equation}
\begin{aligned}
 \frac{\partial D\phi_*}{\partial n_1^*} = \frac{\partial D\phi_*}{\partial p_1^*} = \begin{bmatrix}
    \exp(n_1^* + p_1^*) & \exp(n_1^* + p_1^*) & 0 & 0\\
    \exp(n_1^* + p_1^*) & \exp(n_1^* + p_1^*) & 0 & 0
\end{bmatrix},
\end{aligned}
\end{equation}
and
\begin{equation}
\begin{aligned}
 \frac{\partial D\phi_*}{\partial n_2^*} = \frac{\partial D\phi_*}{\partial p_2^*} = \begin{bmatrix}
    0 & 0 & \exp(n_2^* + p_2^*) & \exp(n_2^* + p_2^*)\\
    0 & 0 & \exp(n_2^* + p_2^*) & \exp(n_2^* + p_2^*)
\end{bmatrix},
\end{aligned}
\end{equation}
we see that a local null vector $v$ satisfies the first-order invariance condition if and only if $v_1+v_2=0$ and $v_3+v_4=0$, since the exponential factors are non-zero. These are exactly the constant null directions, along which $n_1p_1$ and $n_2p_2$ do not change, while a vector in the remaining null direction at one point does not remain a null direction under perturbations of the parameters. In this exact structured transformation example, the condition can thus be evaluated at a single point and recovers the invariant subspace in a purely numerical approach, without needing to solve partial differential equations or perform symbolic calculations. This example therefore illustrates how the invariant image approach can recover an exact image reparameterisation that is not minimal, and how the first-order invariance condition can be used to determine the invariant subspace in a purely numerical approach.

Although not minimal, this reduction is still useful, as it shows that the model depends on the original parameters only through the pair $(n_1p_1,n_2p_2)$. At this point, applying the same log-monomial structure again would only search for monomials of these monomial coordinates, and hence for monomials in the original parameters as before. This does not capture the remaining additive dependence on the sum of the two coordinates. In this example the remaining structure is additive, so we instead use the identity transformation for $f$ and $g$, corresponding to a search for linear combinations of the image coordinates obtained in the first step; in this example, these are linear combinations of monomial coordinates. In practice this structure would not generally be known in advance; the point is that the first-stage image coordinates provide a lower-dimensional space in which further structured reductions, such as additive rather than monomial reductions, can be tested. This could form the basis for a more sophisticated search procedure for finding reparameterisations, but we leave this for future work.

In the present case, we define the variables
\begin{equation}
\begin{aligned}
    \xi = \begin{bmatrix}
    \xi_1\\
    \xi_2
    \end{bmatrix} =
    \begin{bmatrix}
    n_1p_1\\
    n_2p_2
    \end{bmatrix},
\end{aligned}
\end{equation}
as our initial parameterisation. Our auxiliary mapping in these coordinates is given by
\begin{equation}
\begin{aligned}
    \phi_2(\xi_1, \xi_2) =
    \begin{bmatrix}
    \xi_1 + \xi_2\\
    \xi_1 + \xi_2
    \end{bmatrix}.
\end{aligned}
\end{equation}
The Jacobian of this mapping is
\begin{equation}
\begin{aligned}
    D\phi_2(\xi_1, \xi_2) =
    \begin{bmatrix}
    1 & 1\\
    1 & 1
    \end{bmatrix}.
\end{aligned}
\end{equation}
Following the same procedure as above, we can determine the null space and row space of this Jacobian, giving a single identifiable parameter combination $\psi(\xi_1, \xi_2) = \xi_1 + \xi_2 = n_1p_1 + n_2p_2$, and a non-identifiable parameter combination $\lambda(\xi_1, \xi_2) = \xi_1 - \xi_2 = n_1p_1 - n_2p_2$. This is the expected result, and illustrates how the invariant image approach can be applied sequentially to obtain the minimal image reparameterisation. Future work could explore more sophisticated search procedures for finding more general classes of reparameterisations, with the invariant image approach and invariance test used to guide the search.

\subsection{Numerical diagnostic notes}
The numerical non-limit example is structurally identifiable and therefore has no exact invariant null direction. The accompanying Julia implementation prints an additional directional practical-identifiability diagnostic for this case. This diagnostic perturbs the weakest local singular-vector direction in log-parameter space on either side of the maximum-likelihood estimate and recomputes the relative directional sensitivity. These diagnostics are not needed for the main claims of the paper and are not reproduced here; they are available in the repository script \texttt{examples/stat\_model.jl} and its output logs.

\section{Repressilator numerical details}
Here we provide additional details on the repressilator example, including parameter values, observation settings, numerical settings, and diagnostics.

\subsection{Parameter values and observation settings}
The repressilator model has 18 estimated parameters, with the Hill coefficient fixed at $n=2.5$. The state vector $X$ contains the three mRNA and three protein concentrations. The synthetic data use initial condition $X_0=[1,0,0,0,0,0]^T$, 8 equally spaced observation times on $[0,10000]$, and independent Gaussian observation noise with standard deviation $\sigma=10$ on the three mRNA trajectories. The random seed used for the synthetic observations is 42.

\begin{table}[htbp]
\centering
\caption{True parameter values and bounds used for the repressilator example. Estimation bounds were used for the initial maximum-likelihood search. Wider profile bounds were used in the two-dimensional profile-likelihood calculation for the representative $(K_1/\beta_1,\beta_1K_1)$ interest pair.}
\label{tab:supp-repressilator-params}
\begin{tabular}{c|c|c|c}
\hline
Parameter group & True values $(i=1,2,3)$ & Estimation bounds & Profile bounds \\
\hline
$\alpha_{0i}$ & $(0.008,0.009,0.010)$ & $[0.005,0.015]$ & $[0.003,0.020]$ \\
$\alpha_i$ & $(1.0,1.2,1.5)$ & $[0.8,2.0]$ & $[0.5,3.0]$ \\
$\beta_i$ & $(0.020,0.025,0.015)$ & $[0.01,0.03]$ & $[0.0005,2.0]$ \\
$K_i$ & $(30,28,32)$ & $[20,40]$ & $[0.4,1000]$ \\
$k_{\mathrm{degm},i}$ & $(0.006,0.0055,0.0065)$ & $[0.004,0.008]$ & $[0.003,0.010]$ \\
$k_{\mathrm{degp},i}$ & $(0.0012,0.0011,0.0013)$ & $[0.001,0.0015]$ & $[0.0008,0.002]$ \\
\hline
\end{tabular}
\end{table}

For the local IIR calculations used to determine the invariant-image split, and for the augmented-IIR diagnostics discussed below, we solved the repressilator ODE system with the stiff solver \texttt{Rodas4()} on a uniform fine grid of 501 time points on $[0,T_{\mathrm{end}}]$, using absolute and relative tolerances $10^{-10}$ and $10^{-8}$, respectively. Jacobians and the Hessian-based invariance diagnostics were computed by automatic differentiation using \texttt{ForwardDiff.jl}. The numerical rank calculation used relative threshold $\mathrm{rtol}_{\mathrm{rank}}=10^{-7}$, and the invariance test used $\mathrm{rtol}_{\mathrm{invariance}}=10^{-6}$. For sparse monomial basis selection, we searched over integer exponent vectors with support at most two and coefficients in $\{-1,0,1\}$, retaining candidates with projection residual below $10^{-2}$.

\subsection{Sparse basis and augmented observation-map diagnostics}
For the mRNA-only auxiliary map used in the main repressilator analysis, the IIR calculation gives rank 15 and an invariant-null dimension of 3. The sparse null-side basis is
\begin{equation}
\beta_1K_1,\qquad \beta_2K_2,\qquad \beta_3K_3.
\end{equation}
On the identified side, ordering candidates by simplicity, as on the null side, gives a sparse mechanistic basis consisting of the three ratios
\begin{equation}
K_1/\beta_1,\qquad K_2/\beta_2,\qquad K_3/\beta_3,
\end{equation}
together with the remaining 12 singleton directions
\begin{equation}
\alpha_{01},\alpha_{02},\alpha_{03},\alpha_1,\alpha_2,\alpha_3,
 k_{\mathrm{degm},1},k_{\mathrm{degm},2},k_{\mathrm{degm},3},
 k_{\mathrm{degp},1},k_{\mathrm{degp},2},k_{\mathrm{degp},3}.
\end{equation}
The same basis is returned at all eight fits used for the reference-point check in the main text. The information-ordered search also selects these three ratios, but combines the remaining parameters into products and ratios spanning the same invariant image. Depending on sign conventions for exponent vectors, the reciprocal ratios $\beta_i/K_i$ may be returned by the sparse search; these span the same one-dimensional coordinates as $K_i/\beta_i$.

To support the unobserved-protein prediction analysis in the main text, we also ran local IIR diagnostics with augmented auxiliary maps. These use the same reference point and fine time grid as the mRNA-only calculation, but include additional model outputs before computing the Jacobian and invariant null space. The results are summarised in Table~\ref{tab:supp-augmented-iir}. The same holds for each protein: measuring $p_j$ at the observation times removes only the direction $\beta_jK_j$, measuring any two proteins removes two of the three directions, and measurements at $t=0$ remove none, because the protein initial conditions are zero. In the notation of Section~\ref{sec:supp-augmented-maps}, taking $q$ to be $p_j$ at the observation times gives $\rank(Dq\,V_0)=1$, since $p_j$ changes, to first order, along $\beta_jK_j$ but not along the other two null directions.

\begin{table}[htbp]
\centering
\caption{Local augmented-IIR diagnostics for the repressilator. Adding the first protein trajectory removes the $\beta_1K_1$ invariant-null direction while retaining the corresponding directions for genes 2 and 3. Adding all protein trajectories gives full rank for this 18-parameter model.}
\label{tab:supp-augmented-iir}
\begin{tabular}{c|c|c|c}
\hline
Auxiliary map & Rank & Invariant-null dimension & Sparse null labels \\
\hline
$(m_1,m_2,m_3)$ & 15 & 3 & $\beta_1K_1,\beta_2K_2,\beta_3K_3$ \\
$(m_1,m_2,m_3,p_1)$ & 16 & 2 & $\beta_2K_2,\beta_3K_3$ \\
$(m_1,m_2,m_3,p_1,p_2,p_3)$ & 18 & 0 & none \\
\hline
\end{tabular}
\end{table}

\subsection{Full repressilator prediction bands}

\begin{figure}[p]
    \centering
    \includegraphics[width=\linewidth]{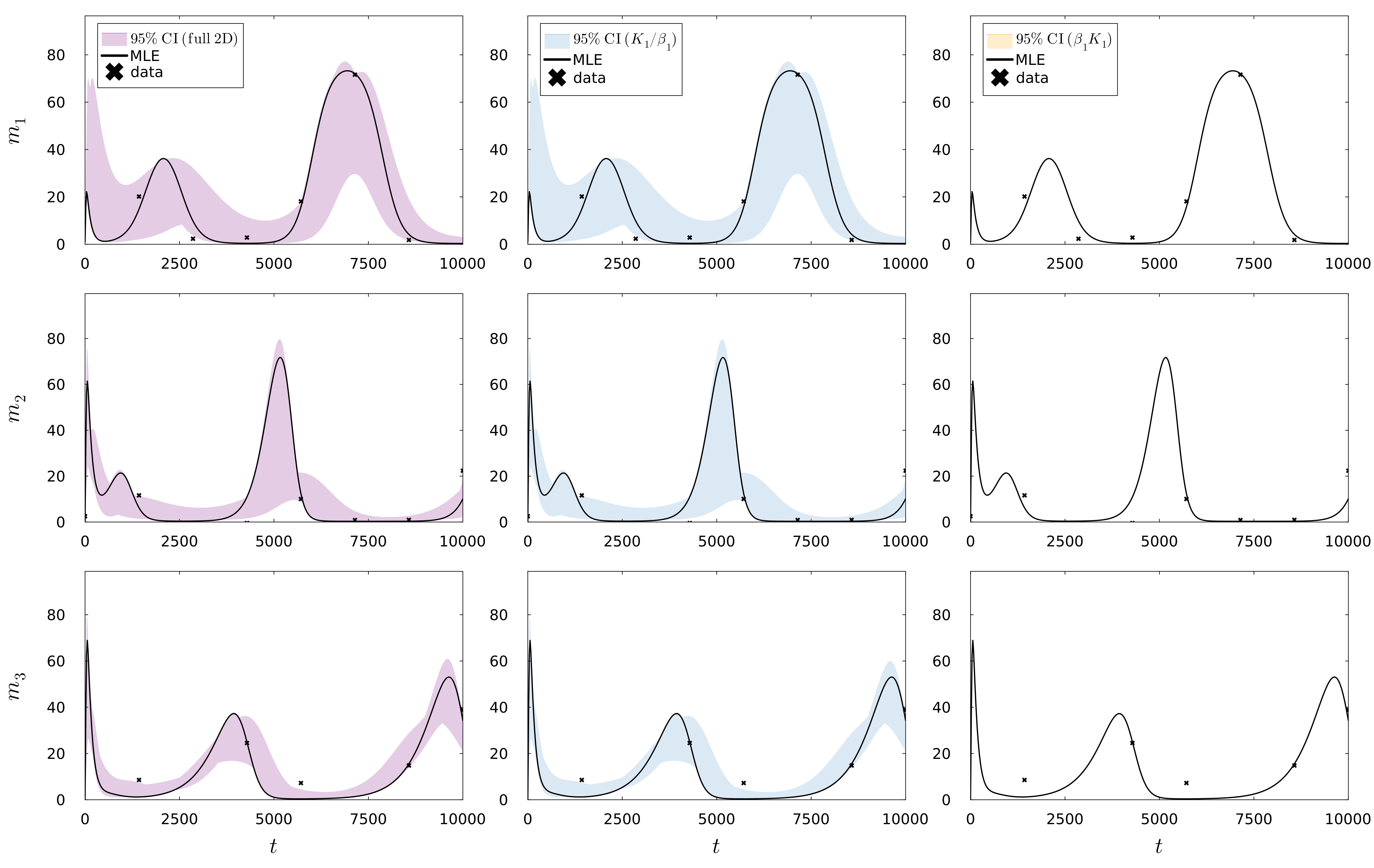}
    \caption{Full profile-wise prediction bands for the observed repressilator mRNA trajectories.}
    \label{fig:supp-repressilator-mrna-predictions}
\end{figure}

\begin{figure}[p]
    \centering
    \includegraphics[width=\linewidth]{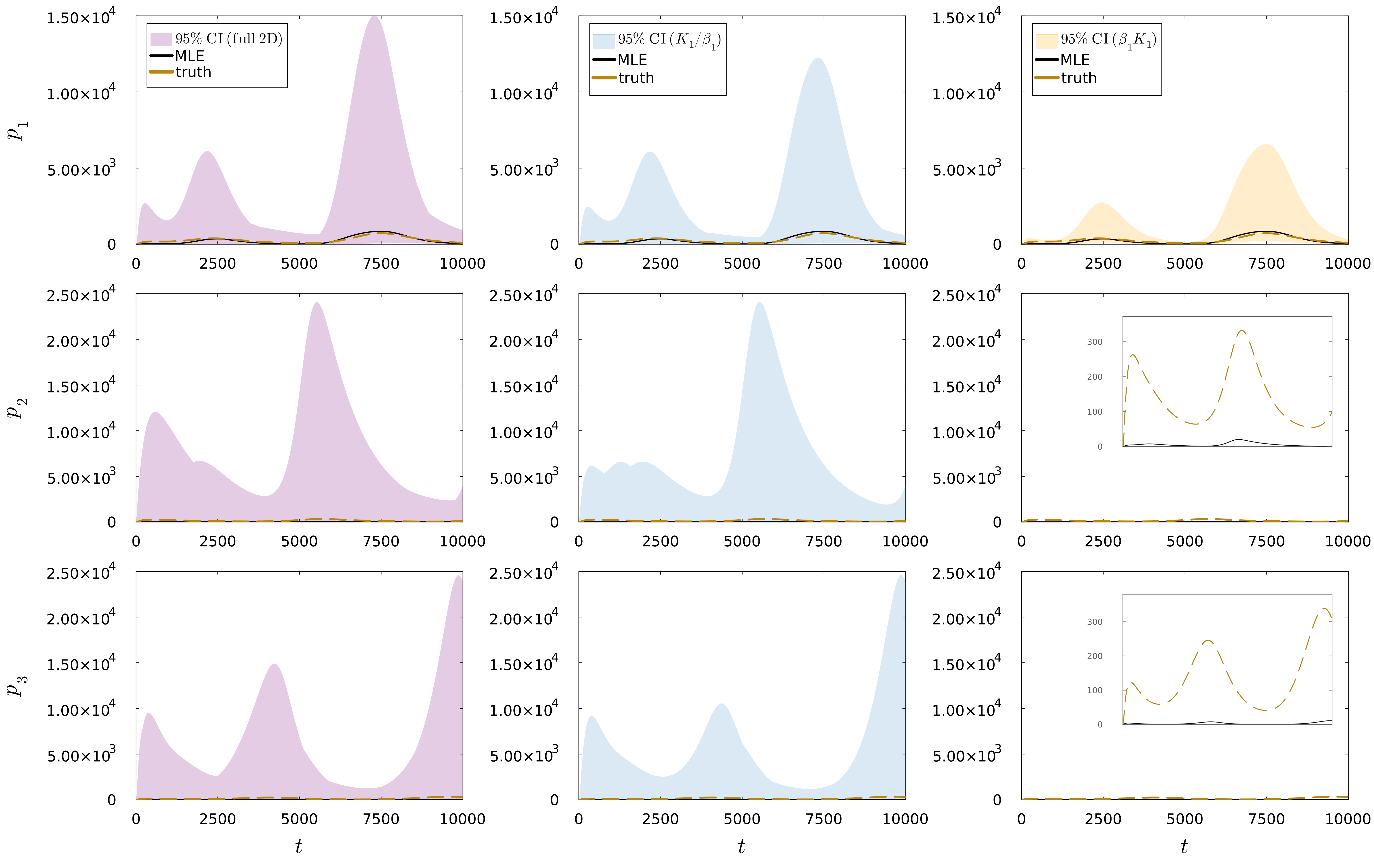}
    \caption{Full profile-wise prediction bands for the unobserved repressilator protein trajectories. Insets in the $p_2$ and $p_3$ panels for the $\beta_1K_1$ profile show the same profiles on a zoomed scale.}
    \label{fig:supp-repressilator-protein-predictions}
\end{figure}

Figure~\ref{fig:supp-repressilator-mrna-predictions} shows the full set of profile-wise prediction bands for the observed mRNA trajectories. Figure~\ref{fig:supp-repressilator-protein-predictions} shows the corresponding bands for the unobserved protein trajectories. The main text shows the representative $m_1$ and $p_1$ rows.

\clearpage
\subsection{Saved results and postprocessing}
The publication figures for the repressilator in the main text were generated by the repository postprocessing scripts from the saved two-dimensional repressilator profile-likelihood results file in the accompanying repository:
\begin{center}
\small\path{nesi/repressilator_16nuisance_50x50_results.jls}
\end{center}
The likelihood-profile figure is produced by \path{replot_profile_results.jl}. The mRNA and protein prediction-band figures are produced by the repository prediction-interval postprocessing script. Prediction bands are generated by pushing forward accepted points from this saved profile surface; no new profile optimisation is performed during this postprocessing.

\section{Additional repository examples}
The accompanying repository contains additional maintained examples that are not used for the main claims of the paper but illustrate the same implementation workflow. The Michaelis--Menten/Monod example demonstrates the contrast between an exact limit case and a full-rank non-limit case with a practically weak direction. The heterogeneous-flow/transport example demonstrates an exact structurally non-identifiable model with a sparse ratio family and illustrates the distinction between orthogonal SVD coordinates and sparse mechanistic coordinates. These examples appeared in a previous version of the manuscript as a preprint but are now available as repository examples rather than full supplementary case studies due to space constraints and to keep the focus on the examples in the main text.

\end{document}